\documentclass[aps,pra,twocolumn,superscriptaddress]{revtex4-1}
\usepackage{amsmath,amssymb,mathtools}
\usepackage[pdftex]{graphicx}
\usepackage[pdftex,colorlinks=true,bookmarks=true,citecolor=blue,linkcolor=blue,urlcolor=blue,breaklinks=true]{hyperref}

\begin{document}

\title{Optomagnonic Barnett effect}

\author{Kouki Nakata}
\affiliation{Advanced Science Research Center, Japan Atomic Energy Agency, 
Tokai, Ibaraki 319-1195, Japan}

\author{Shintaro Takayoshi}
\affiliation{Max Planck Institute for the Physics of Complex Systems,
Dresden 01187, Germany}
\affiliation{Department of Physics, Konan University,
Kobe 658-8501, Japan}

\date{\today}

\begin{abstract}
Combining the technologies of quantum optics and magnonics, 
we find that the circularly polarized laser can dynamically realize 
the quasiequilibrium magnon Bose-Einstein condensates (BEC).
The Zeeman coupling between the laser and spins generates 
the optical Barnett field,
and its direction is controllable by switching the laser chirality.
We show that the optical Barnett field develops the total magnetization 
in insulating ferrimagnets with reversing the local magnetization, 
which leads to the quasiequilibrium magnon BEC.
This laser-induced magnon BEC transition through optical Barnett effect, 
dubbed the optomagnonic Barnett effect,
provides an access to coherent magnons in the high frequency regime 
of the order of terahertz.
We also propose a realistic experimental setup 
to observe the optomagnonic Barnett effect 
using current device and measurement technologies as well as the laser chirping. 
The optomagnonic Barnett effect is a key ingredient 
for the application to ultrafast spin transport. 
\end{abstract}

\maketitle

\section{Introduction}
\label{sec:intro}
For a fast and flexible manipulation of magnetic systems, 
inventing methods to handle magnetism 
is a central task in the field of spintronics. 
Since the seminal works in 1915
by Barnett, Einstein, and de Haas~\cite{Barnett,Barnett2,EdH}, 
the transfer of angular momentum from mechanical rotations 
to spin angular momentum and its reciprocal phenomenon,
dubbed the Barnett effect and the Einstein-de Haas effect, respectively, 
have been intensively investigated. 
Recent progresses are 
the observations of the Barnett effect in paramagnets~\cite{OnoBarnett} 
and in nuclear spin systems~\cite{ChudoBarnett,NuclearBarnett}. 
Another important advance in the manipulation of magnetism 
is the utilization of laser-matter 
coupling~\cite{Mukai2014APL,LaserPhotoExp,Ciappina2017RepProgPhys,LaserPhotoExp3}, 
and the reversal of magnetization is achieved experimentally by means of the optical 
method~\cite{OtherOpticalBarnett,OtherOpticalBarnett2,OtherOpticalBarnett3,Kirilyuk,KimelNatureIFaraday}. 
Thus the interdisciplinary field between optics and 
spintronics~\cite{OptomagnonicsCavity,OptomagnonicsCavity2,OptomagnonicsCavity3,ReviewMagnon,MagnonSpintronics}
attracts a broad interest of both experimentalists and theorists.

The well-known phenomenon for the laser-induced magnetization
is the inverse Faraday effect~\cite{IFE,Kirilyuk,KimelNatureIFaraday}.
The applied laser introduces the coupling to the optical polarization
and induces an emergent effective magnetic field.
The magnitude of the effective field is proportional 
to a square of the laser field. 
Another approach to develop the uniform magnetization 
is to use the Zeeman coupling between 
the circularly polarized laser and 
spin systems~\cite{OtherOpticalBarnett2,OtherOpticalBarnett3,FloquetST2,FloquetST}.
The spin-photon coupling induces an effective magnetic field 
in the direction perpendicular to the laser polarization plane, 
which gives rise to the magnetization. 
Since it is analogous to the generation of magnetization 
by mechanical rotations through spin-rotation coupling, 
i.e., the Barnett effect~\cite{Barnett,Barnett2,SRC,SRC2,SRC3,SRCmm,SRCmm2,SRCmm3,SRCmm4,ReviewSpinMechatronics}, 
the emergence of magnetization through the spin-photon coupling
is dubbed as the optical Barnett effect~\cite{OtherOpticalBarnett2,OtherOpticalBarnett3}.
The effective magnetic field induced by laser, 
an analog of the conventional Barnett field,
is called the optical Barnett field~\cite{OtherOpticalBarnett2,OtherOpticalBarnett3}.
In contrast to the inverse Faraday effect,
the optical Barnett field is independent of the laser field strength,
while it is proportional to the laser frequency~\cite{OtherOpticalBarnett2,OtherOpticalBarnett3,FloquetST2,FloquetST}.

\begin{figure}[t]
\centering
\includegraphics[width=0.47\textwidth]{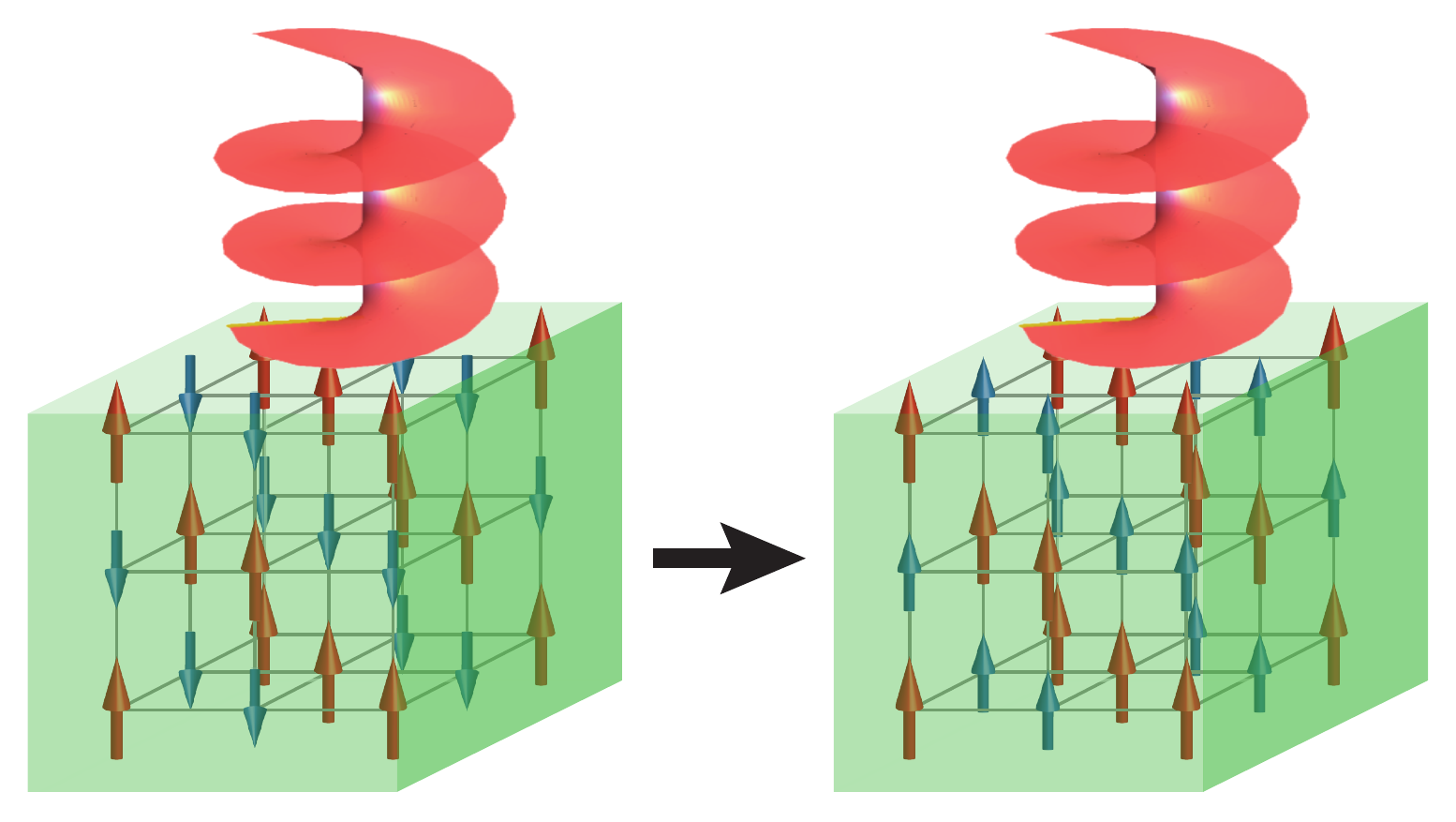}
\caption{
Schematic picture of the optomagnonic Barnett effect
in the insulating ferrimagnet.
}
\label{fig:OpticalBarnett}
\end{figure}

In this paper, 
we investigate an application of circularly polarized laser 
to insulating ferrimagnets, 
following the scheme to introduce a uniform magnetization by 
laser in quantum spin systems~\cite{FloquetST2,FloquetST}. 
We find that the induced optical Barnett field reverses the local magnetization 
and develops the uniform magnetization, which leads to the formation of
the quasiequilibrium magnon Bose-Einstein condensates (BEC).
We give a microscopic description of this magnon BEC transition 
in insulating ferrimagnets. 
We numerically show that the magnetization makes a precession 
with the frequency same as the laser. 
Hence the optical Barnett effect provides an access 
to coherent magnons in the high frequency regime of the order of terahertz.
Since this result arises from the combination of 
quantum optics and magnon spintronics (i.e., magnonics),
we refer to this optical Barnett effect 
especially as the optomagnonic Barnett effect. 
Thus the optomagnonic Barnett effect enables us 
to control magnons coherently in much faster time scale than 
the conventional microwave pumping.
We also propose a realistic experimental setup 
using ferrimagnetic insulators and the chirping technique 
of circularly polarized laser. 
Our findings play a role of building blocks 
for the application to ultrafast spin transport.

This paper is organized as follows.
In Sec.~\ref{sec:OBeffect}
we quickly review the mechanism of the optical Barnett effect,
and find the optomagnonic Barnett effect in Sec.~\ref{sec:optomagnonicBE}. 
In Sec.~\ref{sec:Experiment},
we discuss the experimental feasibility.
Finally, we remark on several issues in Sec.~\ref{sec:Discussion}
and summarize in Sec.~\ref{sec:conclusion}.
Technical details are described in the Appendices.

\begin{table}[t]
\centering
\caption{Comparison between the mechanical and optical Barnett effects.
}
\begin{tabular}{c|cc}
\hline
 & Mechanical Barnett & Optical Barnett \\
\hline
Induced by & Mechanical rotation & Circularly polarized laser \\
Coupling & Spin-rotation & Spin-photon \\
Barnett field  & $\propto$ Angular velocity & 
$\propto$ Laser frequency \\
\hline
\end{tabular}
\label{tab:Compare}
\end{table}

\section{Optical Barnett effect}
\label{sec:OBeffect}

In this section, we quickly review the mechanism that 
the Zeeman coupling between circularly polarized laser and spins 
induce an effective magnetic field perpendicular to 
the laser polarization plane, 
which develops the uniform magnetization~\cite{FloquetST2,FloquetST}. 
We explain the analogy between 
this phenomenon, the optical Barnett effect, 
and the Barnett effect caused by mechanical rotations. 
Hereafter we use the terminology \textit{mechanical} Barnett effect (field) 
to mean the conventional Barnett effect (field) by the mechanical rotation
in order to distinguish it from the optical one. 
The comparison between the optical and mechanical 
Barnett effects is summarized in Table~\ref{tab:Compare}.

Let us consider quantum spin systems described by the Hamiltonian 
$\mathcal{H}_{0}$. We take the polarization plane as the $xy$ plane 
and the $z$ axis as the direction perpendicular to it. 
We assume that $\mathcal{H}_{0}$ has the $U(1)$ symmetry about the $z$ 
axis for simplicity. Here we focus on the magnetic insulator 
with a large electronic gap, and only consider the Zeeman coupling 
between the spins and magnetic component of laser. 
The time-periodic Hamiltonian is written as~\cite{FloquetST2,FloquetST}
\begin{align}
 \mathcal{H}(t)=\mathcal{H}_{0}
   -B_{0}[S_{\mathrm{tot}}^{x}\cos(\Omega t)
     +\eta S_{\mathrm{tot}}^{y}\sin(\Omega t)],
\label{eq:Hamil_tdep}
\end{align}
where $B_{0}>0$ and $\Omega>0$ are respectively 
the magnetic field amplitude and the frequency, 
i.e., photon energy, of the laser. 
The sign $\eta=+(-)$ represents the left (right) circular polarization, 
and $S_{\mathrm{tot}}^{x(y,z)} \coloneqq \sum_{j}S_{j}^{x(y,z)}$ 
is the summation over spin operators on all the spin sites. 
Through the Floquet theory or the unitary transformation 
$\mathcal{H}(t)\to e^{i\eta \Omega t S_{\mathrm{tot}}^{z}}
(\mathcal{H}(t)-i\hbar\partial_{t})e^{-i\eta \Omega t S_{\mathrm{tot}}^{z}}$,
we derive an effective static Hamiltonian~\cite{FloquetST2,FloquetST} (Appendix~\ref{sec:effHamil}) 
\begin{align}
 \mathcal{H}_{\mathrm{eff}}
   =\mathcal{H}_{0}-\eta\hbar\Omega S_{\mathrm{tot}}^{z}
   +O(B_{0}).
\label{eq:Hamil_eff}
\end{align}
Here we consider the case of weak laser field 
$B_{0}\ll \hbar\Omega$, 
and the $B_{0}S_{\mathrm{tot}}^{x}$ term is negligibly small.
From Eq.~\eqref{eq:Hamil_eff}, we see that the circularly polarized laser 
introduces the effective coupling $-\eta\hbar\Omega S_{\mathrm{tot}}^{z}$, 
which plays the same role as 
the mechanical Barnett field~\cite{SRC,SRC2,SRC3,SRCmm,SRCmm2,SRCmm3,SRCmm4,ReviewSpinMechatronics} 
obtained from the spin-rotation coupling (Table~\ref{tab:Compare}). 
This effective coupling is recast into the Zeeman-type interaction  
$-\eta\hbar\Omega S_{\mathrm{tot}}^{z}= -\eta\hbar\gamma\mathcal{B}S_{\mathrm{tot}}^{z}$ 
with the gyromagnetic ratio $\gamma$ 
and we refer to
\begin{align}
 \mathcal{B} \coloneqq \Omega/\gamma
\end{align}
as the \textit{optical Barnett field}~\cite{OtherOpticalBarnett2,OtherOpticalBarnett3}. 
This optical Barnett field develops the total magnetization 
and plays an essential role in the optical Barnett effect. 
The direction of the optical Barnett field is controllable through the change of 
the laser chirality, i.e., circular polarization, $\eta=\pm$~\cite{FloquetST2,FloquetST}.

We remark that Eq.~\eqref{eq:Hamil_eff} holds for a general $U(1)$ symmetric 
spin Hamiltonian $\mathcal{H}_{0}$, which indicates that essentially 
any kind of magnets, e.g., electron and nuclear spin systems, even paramagnets, 
can exhibit the optical Barnett effect.
Moreover, the induced term $\eta\hbar\Omega S_{\mathrm{tot}}^{z}$ 
is independent of material parameters such as $g$ factor, 
and only depends on the laser parameters. 
In that sense, we can say that the optical Barnett effect 
is a \textit{universal} phenomenon. 
Note that the circularly polarization is the key ingredient 
of the optical Barnett effect.
Since the \textit{linearly} polarized laser does not 
develop magnetization~\cite{FloquetST2}, 
it neither produces the optical Barnett field.

While we treat the laser as a classical electromagnetic field in the above, 
we can explain the same phenomenon through the spin-photon coupling. 
Since the photon has spin $\pm 1$ depending on 
the circular polarization of laser $\eta=\pm$, 
the Hamiltonian is given as 
$\mathcal{H}=\mathcal{H}_{0}-g_{\mathrm{s}\mathchar`-\mathrm{ph}}
\sum_{j}(a_{j}S_{j}^{\eta}+a_{j}^{\dagger} S_{j}^{-\eta})
+\hbar\Omega\sum_{j}a_{j}^{\dagger}a_{j}$, 
where $S^{\pm} \coloneqq S^{x}\pm iS^{y}$, 
$a^{\dagger}$ and $a$ are the bosonic creation and annihilation operators of photons, 
and $g_{\mathrm{s}\mathchar`-\mathrm{ph}}$ is the spin-photon coupling constant, 
which is proportional to $B_{0}$. 
Noting that the total spin angular momentum 
$\eta \sum_{j}a_{j}^{\dagger}a_{j}+S_{\mathrm{tot}}^{z}$ 
is conserved, we substitute 
$\sum_{j}a_{j}^{\dagger}a_{j}=\mathrm{const.}-\eta S_{\mathrm{tot}}^{z}$ 
into the Hamiltonian and obtain 
$\mathcal{H}=\mathcal{H}_{0}-\eta\hbar\Omega S_{\mathrm{tot}}^{z}-g_{\mathrm{s}\mathchar`-\mathrm{ph}}
\sum_{j}(a_{j}S_{j}^{\eta}+a_{j}^{\dagger} S_{j}^{-\eta})$. 
In the case of $B_{0}\ll \hbar\Omega$, this Hamiltonian coincides with 
Eq.~\eqref{eq:Hamil_eff}. 
Thus the spin angular momentum of photon is transferred to 
the magnet in the optical Barnett effect, 
and we can understand it analogously with 
the mechanical Barnett effect (Table~\ref{tab:Compare}).

\begin{figure}[t]
\centering
\includegraphics[width=0.47\textwidth]{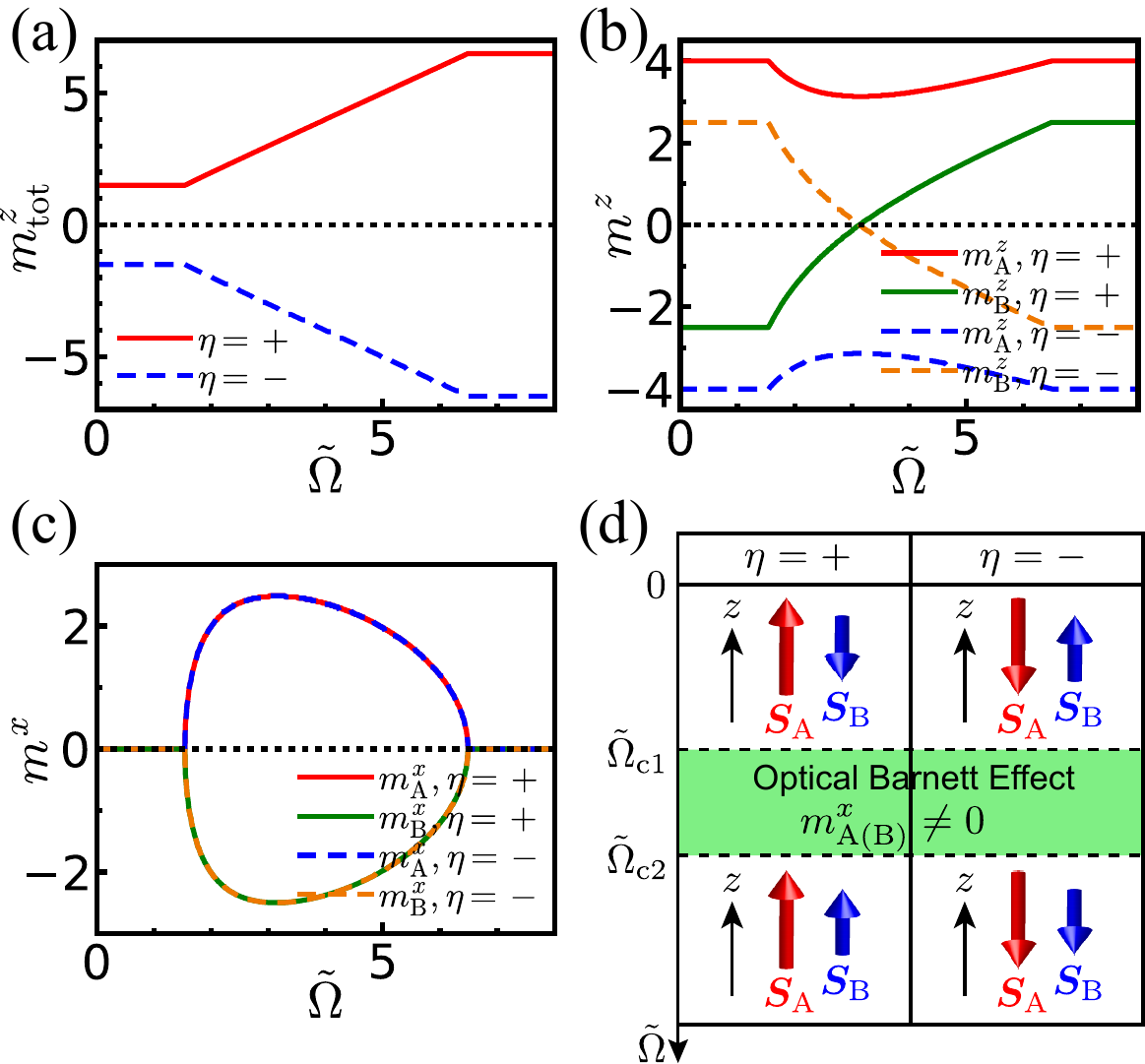}
\caption{The classical spin configuration that minimizes 
the energy $\epsilon$ [Eq.~\eqref{eq:Classical_E}]. 
(a) $m_{\mathrm{tot}}^{z}$, (b) $m_{\mathrm{A(B)}}^{z}$, and 
(c) $m_{\mathrm{A(B)}}^{x}$ are shown as a function of 
$\tilde{\Omega}=\hbar \Omega/(z_{0}J)$. 
The direction of the optical Barnett field depends 
on the laser chirality $\eta$. 
The parameters are $S_{\mathrm{A}}=4$, $S_{\mathrm{B}}=5/2$, 
$D_{\mathrm{A}}/J=17.5\times 10^{-3}$, and 
$D_{\mathrm{B}}/J=1.5\times 10^{-3}$ 
following the experimental values for 
$\mathrm{Er_{3}Fe_{5}O_{12}}$~\cite{ohnuma,FerriOhnuma,FerriOhnuma2} 
($J=0.2$ meV).
The lower and upper critical frequencies are 
$\tilde{\Omega}_{\mathrm{c1}}=1.54$ and
$\tilde{\Omega}_{\mathrm{c2}}=6.49$, respectively.
(d) Summary of the optical Barnett effect and the magnetization reversal 
on the sublattice B for both circular polarization of laser.
}
\label{fig:magnetization}
\end{figure}

\section{Optomagnonic Barnett effect}
\label{sec:optomagnonicBE}

In this paper we discuss 
the formation of
the quasiequilibrium magnon BEC provoked by
the optical Barnett effect,
which we call the optomagnonic Barnett effect.
As a platform, 
we consider the laser application to insulating ferrimagnets (Fig.~\ref{fig:OpticalBarnett}), 
\begin{align}
 \mathcal{H}_{0}
   =&J\sum_{\langle i\in \mathrm{A}, j\in \mathrm{B} \rangle}
     \boldsymbol{S}_{\mathrm{A},i} \cdot
     \boldsymbol{S}_{\mathrm{B},j}\nonumber\\
     &-D_{\mathrm{A}}\sum_{i\in\mathrm{A}}
       (S_{\mathrm{A},i}^{z})^{2}
     -D_{\mathrm{B}}\sum_{j\in\mathrm{B}}
       (S_{\mathrm{B},j}^{z})^{2},
\label{eq:Hamil_Ferri}
\end{align}
where $\boldsymbol{S}_{\mathrm{A(B)}, i(j)}=(S_{\mathrm{A(B)}, i(j)}^{x}, 
S_{\mathrm{A(B)}, i(j)}^{y}, S_{\mathrm{A(B)}, i(j)}^{z})$ 
represents the spin at the $i(j)$-th site on the sublattice $\mathrm{A(B)}$ 
having the spin quantum number $S_{\mathrm{A(B)}}$,
$J>0$ is the exchange interaction between the nearest neighbor spins 
$\langle i\in \mathrm{A}, j\in \mathrm{B} \rangle$,
and $D_{\mathrm{A(B)}}>0$ is the easy-axis single ion anisotropy for the sublattice $\mathrm{A(B)}$ 
that ensures a magnetic order in the $z$ direction. 
In the systems with anisotropy, 
we can realize the dynamical magnetization curve 
by modulating the laser frequency $\Omega$ slowly enough~\cite{FloquetST}, 
which is the experimental technique called chirping~\cite{chirping,chirping2}. 

We remark that in antiferromagnets
($S_{\mathrm{A}}=S_{\mathrm{B}}$) with easy-axis anisotropy, 
the spin-flop transition happens 
in the low field regime associated with the N\'eel magnetic order 
when the static external field is increased~\cite{SpinFlopAF}. 
The spin-flop transition is of the first order 
and the change of the state is drastic. 
In the case of laser application, 
the dynamical state cannot follow this sudden change, 
and the optical Barnett effect does not take place. 
In ferrimagnets ($S_{\mathrm{A}}\not=S_{\mathrm{B}}$), however, 
the spin-flop transition is absent~\cite{SpinFlopTra},
and that is why we consider ferrimagnets in this paper.

\subsection{Classical theory}
\label{subsec:OMBEC1}

First we analyze the optical Barnett effect in the classical case. 
Since the effective Hamiltonian Eq.~\eqref{eq:Hamil_eff}, 
where $\mathcal{H}_{0}$ is Eq.~\eqref{eq:Hamil_Ferri}, 
has the $U(1)$ symmetry, 
we assume that the spins reside in the $xz$ plane, 
$\boldsymbol{S}_{\mathrm{A},i}=(m_{\mathrm{A}}^{x}, 0, m_{\mathrm{A}}^{z})$ and 
$\boldsymbol{S}_{\mathrm{B},j}=(m_{\mathrm{B}}^{x}, 0, m_{\mathrm{B}}^{z})$. 
The classical energy normalized by the number of spins is given as 
\begin{align}
 \epsilon
   =\frac{z_{0}J}{2}\boldsymbol{S}_{\mathrm{A}}\cdot\boldsymbol{S}_{\mathrm{B}}
     -\frac{D_{\mathrm{A}}}{2}(S_{\mathrm{A}}^{z})^{2}
     -\frac{D_{\mathrm{B}}}{2}(S_{\mathrm{B}}^{z})^{2}
     -\eta \frac{\hbar\Omega}{2}(S_{\mathrm{A}}^{z}+S_{\mathrm{B}}^{z}),
\label{eq:Classical_E}
\end{align}
where $z_{0}$ is the coordination number. 
We numerically obtain the classical spin configuration 
that minimizes the energy [Eq.~\eqref{eq:Classical_E}], 
and show the result in Fig.~\ref{fig:magnetization}. 
Here we consider the cubic lattice with $z_{0}=6$. 
The magnetization curve 
induced by the optical Barnett field, i.e., 
$m_{\mathrm{tot}}^{z}\coloneqq m_{\mathrm{A}}^{z}+m_{\mathrm{B}}^{z}$ 
as a function of the normalized frequency 
$\tilde{\Omega}\coloneqq \hbar\Omega/(z_{0}J)$, 
is shown in Fig.~\ref{fig:magnetization}(a). 
When the frequency $\tilde{\Omega}$ is small, the spin configuration is unchanged and 
aligned along the $z$ direction due to the anisotropy. 
Above the lower critical frequency $\tilde{\Omega}_{\mathrm{c1}}$, 
the total magnetization along the $z$ axis starts to grow. 
In this optical Barnett effect, 
$m_{\mathrm{tot}}^{z}$ increases continuously and attains full polarization 
at the upper critical frequency $\tilde{\Omega}_{\mathrm{c2}}$. 
Fig.~\ref{fig:magnetization}(b) shows the change of 
$m_{\mathrm{A}}^{z}$ and $m_{\mathrm{B}}^{z}$ with increasing $\Omega$. 
This indicates that the spins on the sublattice B are reversed 
from $-\eta S_{\mathrm{B}}$ to $\eta S_{\mathrm{B}}$. 
The controllability for the direction of the optical Barnett field by 
the laser chirality $\eta=\pm$ provides a handle to design 
optomagnonic functionalities in various magnets,
e.g., electron and nuclear spin systems, even paramagnets. 
From Fig.~\ref{fig:magnetization}(c), 
we see that both $m_{\mathrm{A}}^{x}$ and $m_{\mathrm{B}}^{x}$ change continuously 
and take nonzero value in 
$\tilde{\Omega}_{\mathrm{c1}}<\tilde{\Omega}<\tilde{\Omega}_{\mathrm{c2}}$. 

Those results of the optical Barnett effect and 
the magnetization reversal in the insulating ferrimagnet
are summarized in Fig.~\ref{fig:magnetization}(d).
The explicit form of $\tilde{\Omega}_{\mathrm{c1(c2)}}$
is given in the Appendix~\ref{sec:classical}.

\subsection{Spin wave theory}
\label{subsec:OMBEC2}

The absence of the first order transition, 
i.e., jump of $m_{\mathrm{tot}}^{z}$, 
in the vicinity of $\tilde{\Omega}_{\mathrm{c1}}$ 
and $\tilde{\Omega}_{\mathrm{c2}}$ ensures the validity of 
the description in terms of the magnon picture. 
Hence we move to the analysis by the spin wave theory next, 
and see that $\tilde{\Omega}_{\mathrm{c1}}$ 
and $\tilde{\Omega}_{\mathrm{c2}}$ become 
the magnon BEC transition points. 

We first consider increasing the frequency $\Omega$ 
from below $\tilde{\Omega}_{\mathrm{c1}}$, 
where the ground state has an alternating structure 
of up and down spins 
[Figs.~\ref{fig:magnetization}(b) and \ref{fig:magnetization}(d)]. 
From the spin wave theory, 
elementary excitations are two kinds of magnons~\cite{ohnuma,KSJD} 
designated by the index $\sigma=\pm$ 
having the spin angular momentum 
$\delta S^{z}=-\eta\sigma 1$. 
The Hamiltonian [Eqs.~\eqref{eq:Hamil_eff} and~\eqref{eq:Hamil_Ferri}]
can be recast into the diagonal form 
due to the $U(1)$ symmetry as 
\begin{align}
 {\mathcal{H}}_{\mathrm{eff}}=\sum_{\sigma=\pm,\boldsymbol{k}}
   (\hbar\omega_{\sigma,\boldsymbol{k}}^{[\alpha]}
     +\Delta_{\sigma}^{[\alpha]}
     +\sigma\hbar\Omega)
   \alpha_{\sigma,\boldsymbol{k}}^{\dagger}\alpha_{\sigma,\boldsymbol{k}},
\label{eq:Hamil_MagBEC1}
\end{align}
where $\Delta_{\sigma}^{[\alpha]}+\sigma\hbar\Omega$ is the magnon gap in laser 
and $\hbar\omega_{\sigma,\boldsymbol{k}}^{[\alpha]}$ 
is the energy dispersion of the $\sigma$ magnon 
annihilated (created) by the bosonic operator 
$\alpha_{\sigma,\boldsymbol{k}}^{(\dagger)}$
with $ [\alpha_{\sigma,\boldsymbol{k}}, \alpha_{\sigma^{\prime},\boldsymbol{k^{\prime}}}^{\dagger} ]
=\delta _{\sigma, \sigma^{\prime}} \delta _{{\boldsymbol{k}}, \boldsymbol{k^{\prime}}} $.
For the details of the calculation and the explicit forms of 
$\Delta_{\sigma}^{[\alpha]}$ and 
$\omega_{\sigma,\boldsymbol{k}}^{[\alpha]}$, 
see the Appendix~\ref{sec:spinwave}. 
With increasing $\Omega$, the energy band of $\sigma=-$ magnon goes down, 
while that of $\sigma=+$ magnon goes up due to the $\sigma\hbar\Omega$ term.
The former touches the zero energy at 
\begin{align}
 \Omega_{\mathrm{BEC1}}
   \coloneqq \Delta_{-}^{[\alpha]}/\hbar,
\end{align}
and the second order phase transition happens 
from the proliferation of magnons.
This is the quasiequilibrium magnon BEC 
induced by the optical Barnett field,
which we call the \textit{optical magnon BEC}. 
$\Omega_{\mathrm{BEC1}}$ coincides with $\Omega_{\mathrm{c1}}$. 
This optical magnon BEC is the macroscopic coherent state 
with the transverse magnetization 
associated with the spontaneous $U(1)$ symmetry 
breaking~\footnote{The total number of magnons in the system 
is bounded by a hard-core interaction 
between magnons~\cite{oshikawa,TotsukaBEC2,Giamarchi2008NatPhys} 
arising from the higher order term in the spin wave theory. 
We neglect it for simplicity in Eqs.~\eqref{eq:Hamil_MagBEC1} 
and \eqref{eq:Hamil_MagBEC2}. 
Thereby the magnon BEC is stable in the system with a finite spin length.},
and thus the total magnetization along the $z$ axis 
grows (Fig.~\ref{fig:magnetization}).
Therefore this optical Barnett effect can be observed as 
the phenomenon induced by the optical magnon BEC transition,
and we refer to this behavior in insulating ferrimagnets
especially as the \textit{optomagnonic Barnett effect}. 

Next we consider decreasing the frequency $\Omega$ 
from above $\tilde{\Omega}_{\mathrm{c2}}$, 
where spins are full polarized in the ground state 
[Figs.~\ref{fig:magnetization}(b) and \ref{fig:magnetization}(d)]. 
Again there are two kinds of magnons designated 
by the index $\sigma=\pm$ 
due to $S_{\mathrm{A}}\not=S_{\mathrm{B}}$, 
but in contrast to the $\Omega_{\mathrm{BEC1}}$ case, 
both magnons have the same spin angular momentum 
$\delta S^{z}=-\eta 1$
since spins on both sublattices are polarized in the same direction. 
We can derive the Hamiltonian in the diagonal form 
\begin{align}
 {\mathcal{H}}_{\mathrm{eff}}=\sum_{\sigma=\pm,\boldsymbol{k}}
   (\hbar\omega_{\sigma,\boldsymbol{k}}^{[\beta]}
     +\Delta_{\sigma}^{[\beta]}
     +\hbar\Omega)
   \beta_{\sigma,\boldsymbol{k}}^{\dagger}\beta_{\sigma,\boldsymbol{k}},
\label{eq:Hamil_MagBEC2}
\end{align}
where $\Delta_{\sigma}^{[\beta]} +\hbar\Omega$ is the magnon gap in laser and 
$\hbar\omega_{\sigma,\boldsymbol{k}}^{[\beta]}$ 
is the energy dispersion of the $\sigma$ magnon 
annihilated (created) by the bosonic operator 
$\beta_{\sigma,\boldsymbol{k}}^{(\dagger)}$
with $ [\beta _{\sigma,\boldsymbol{k}}, \beta _{\sigma^{\prime},\boldsymbol{k^{\prime}}}^{\dagger} ]
=\delta _{\sigma, \sigma^{\prime}} \delta _{{\boldsymbol{k}}, \boldsymbol{k^{\prime}}} $.
For the explicit forms of $\Delta_{\sigma}^{[\beta]} (\leq 0)$ and 
$\omega_{\sigma,\boldsymbol{k}}^{[\beta]} $, 
see the Appendix~\ref{sec:spinwave}. 
With decreasing $\Omega$, the energy band of 
both $\sigma=\pm$ magnon goes down 
due to the $\hbar\Omega$ term, 
and the lower band touches the zero energy at 
\begin{align}
 \Omega_{\mathrm{BEC2}}
   \coloneqq -\Delta_{-}^{[\beta]}/\hbar.
\end{align}
In the same way as the $\Omega_{\mathrm{BEC1}}$ case,
the second order phase transition happens at $\Omega_{\mathrm{BEC2}}$ 
and magnons form the quasiequilibrium BEC. 
$\Omega_{\mathrm{BEC2}}$ coincides with $\Omega_{\mathrm{c2}}$. 
Thus the optomagnonic Barnett effect is induced in the regime 
$\Omega_{\mathrm{BEC1}}<\Omega<\Omega_{\mathrm{BEC2}}$. 

\begin{figure}[t]
\centering
\includegraphics[width=0.47\textwidth]{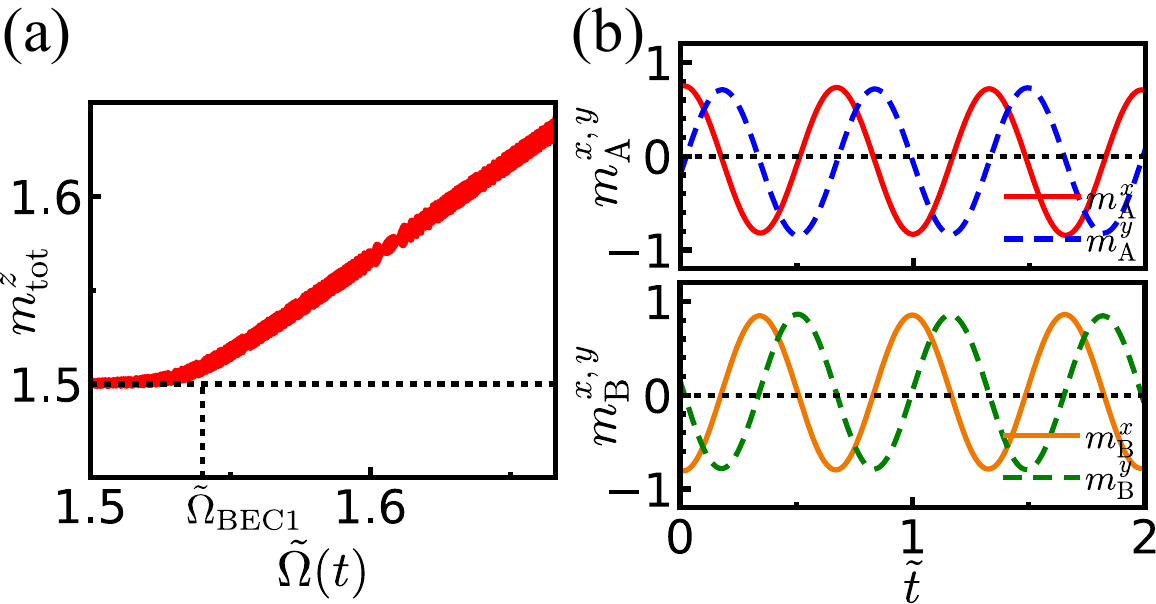}
\caption{The magnetization dynamics for the laser chirality $\eta =+$ 
calculated by the TDMF theory.
Time evolution of 
(a) $m_{\mathrm{tot}}^{z}$ and 
(b) $xy$ components of sublattice magnetization
are shown. The result clearly shows the condensation of magnons 
and the precession of magnetization around the $z$ axis. 
}
\label{fig:dynamics}
\end{figure}

\subsection{Magnetization dynamics}
\label{subsec:OMBEC3}

Finally, to investigate the dynamics of the optomagnonic Barnett effect, 
we numerically solve the equation of motion derived 
from the time-dependent mean field (TDMF) theory 
and calculate the time evolution of sublattice magnetization 
(Appendix~\ref{sec:tdMFT}). 
The TDMF theory can well capture the magnetization dynamics~\cite{Magdynam}. 
The parameters are the same in Fig.~\ref{fig:magnetization}, 
$S_{\mathrm{A}}=4$, $S_{\mathrm{B}}=5/2$, $z_{0}=6$, 
$D_{\mathrm{A}}/J=17.5\times 10^{-3}$, 
and $D_{\mathrm{B}}/J=1.5\times 10^{-3}$.
We use the laser with polarization $\eta=+$ 
represented as $B_{0}(\cos\vartheta(t),\sin\vartheta(t),0)$, 
where the amplitude is $B_{0}/J=0.2$
and the frequency is chirped as 
$\vartheta(t)=\Omega_{0}t+vt^{2}/2$ 
with the normalized chirping speed $\hbar^{2}v/J^{2}=10^{-5}$
and $\hbar\Omega_{0}/J=9$.
The normalized instantaneous frequency is defined as 
$\tilde{\Omega}(t)\coloneqq \hbar (d\vartheta(t)/dt)/(z_{0}J)=(\hbar \Omega_{0}+\hbar vt)/(z_{0}J)$.
We calculate the dynamics in the time region
$0\leq tJ/\hbar\leq 10^{5}$,
which corresponds to
$1.5\leq \tilde{\Omega}(t)\leq 1.66$
in the frequency regime.  
Figure~\ref{fig:dynamics}(a) shows the time evolution of 
$m_{\mathrm{tot}}^{z}=m_{\mathrm{A}}^{z}+m_{\mathrm{B}}^{z}$.
We can see that $m_{\mathrm{tot}}^{z}$ starts to grow 
from $S_{\mathrm{A}}-S_{\mathrm{B}}=3/2$ 
when $\tilde{\Omega}(t)$ exceeds $\tilde{\Omega}_{\mathrm{BEC1}}=1.54$. 
In Fig.~\ref{fig:dynamics}(b), 
we show the time evolution for the $xy$ components 
of sublattice magnetization around $\tilde{\Omega}(t)=1.6 $ 
in the time interval of $0\leq \tilde{t}\leq 2$, 
where $\tilde{t}\coloneqq tJ/\hbar-6\times 10^{4}$ is the normalized time. 
The result clearly shows that magnetization on both sublattices 
precesses around the $z$ axis 
with the instantaneous frequency,
same as the laser $\tilde{\Omega}(t)$, 
and the $xy$ components of 
A and B sublattice magnetization are in the opposite direction. 
The period of this spin precession is $O(1)$ ps.

\section{Experimental feasibility}
\label{sec:Experiment}
We make an estimate for an insulating ferrimagnet 
$\mathrm{Er_{3}Fe_{5}O_{12}}$~\cite{ohnuma,FerriOhnuma,FerriOhnuma2}, 
and give the magnetization curve and the experimental parameter values 
in Fig.~\ref{fig:magnetization} and its caption, respectively.
We find that
the magnon BEC transition points are
$\Omega_{\mathrm{BEC1}}=1.85$ THz and
$\Omega_{\mathrm{BEC2}}=7.8$ THz~\footnote{
Note that the quasiequilibrium magnon BEC reported in Ref.~\cite{demokritov}
is experimentally realized by magnon injection
through microwave pumping in the GHz regime.},
and the optical Barnett field amounts to 
$\mathcal{B}=O(10)$ T for $\Omega=O(1)$ THz.
Our proposal is within the experimental reach 
with current device and measurement technologies, 
e.g., nuclear magnetic resonance~\cite{ChudoBarnett,NuclearBarnett,LeeHahn}
for the optical Barnett field, 
magneto-optical Kerr effect~\cite{KerrEffect}
for the magnetization reversal in the optical Barnett effect, 
Brillouin light scattering~\cite{demokritov} 
for the optical magnon BEC,
and terahertz spectroscopy~\cite{THzspectroscopy,THzspectroscopy2} 
for the spin dynamics of the order of picoseconds.
Since magnons are induced by laser and not by thermal fluctuation 
in the present setup, 
our findings are realizable at low temperature~\cite{magnon10mK,tabuchi,tabuchiScience,Tmagnonphonon}
where phonon degrees of freedom cease to work.

We emphasize the importance of modulating the laser frequency adiabatically~\cite{FloquetST2,FloquetST}
by the chirping technique~\cite{chirping,chirping2}. 
Otherwise, the deviation from the magnetization curve happens 
due to a nonadiabatic transition 
from the Landau-Zener tunneling~\cite{LZ,LZ2}. 
To avoid this effect, a large magnetic anisotropy and 
a strong laser field is advantageous~\cite{FloquetST2}. 
In addition, the laser chirping suppresses heating effects drastically.

\section{Discussion}
\label{sec:Discussion}

First,
the laser application without chirping can be studied 
by the Floquet theory 
with inverse frequency expansion~\cite{FloquetReview,STO2016,KSS}. 
This analysis also supports the generation of 
the optical Barnett field in the high frequency regime 
(Appendix~\ref{sec:highfreq}). 

Second, we remark that
the optical Barnett field through the chirping is proportional to 
the laser frequency $\mathcal{B}\propto\Omega$~\cite{FloquetST2,FloquetST}. 
Hence $\Omega=O(1)$ THz amounts to $\mathcal{B}=O(10)$ T, 
which provides a platform 
to explore the phenomena at high magnetic field
$O(10)$ T or more in the tabletop setup. 

Third, optomagnonic cavities for implementing coherent photon-magnon coupling 
have been theoretically studied in Refs.~\cite{Greece2,Greece4,Greece5,Greece,Greece3}.

Last, as an application of the optomagnonic Barnett effect,
it will be intriguing to investigate the magnon Josephson effect in a junction~\cite{KKPD,Troncoso}.
We leave it for a future study.

\section{Conclusion}
\label{sec:conclusion}

We applied the optical Barnett effect 
to insulating ferrimagnets 
and showed that quasiequilibrium magnon BEC 
can be realized using the spin-wave theory. 
This optomagnonic Barnett effect 
provides an access to coherent magnons 
in the frequency regime of the order of terahertz, 
which is much faster time scale than the conventional microwave pumping.
Our findings are expected to become a building block 
for the application to ultrafast spin transport.

\acknowledgements
The authors would like to thank H. Chudo, Y. Ohnuma, K. Usami, and K. Totsuka 
for useful discussions.
The authors are grateful also to Makoto Oka 
for helpful feedback on this work. 
KN is supported 
by JSPS KAKENHI Grant Number JP20K14420
and 
by Leading Initiative for Excellent Young Researchers, MEXT, Japan.
KN is grateful to the hospitality of MPI-PKS during his stay financially supported by ASRC-JAEA, 
where this work was initiated.

\appendix
\begin{widetext}

\section{Effective static Hamiltonian}
\label{sec:effHamil}

In this section
starting from
the time-periodic Hamiltonian 
\begin{align}
 \mathcal{H}(t)=\mathcal{H}_{0}
   -B_{0}[S_{\mathrm{tot}}^{x}\cos(\Omega t)
     +\eta S_{\mathrm{tot}}^{y}\sin(\Omega t)],
\end{align}
we derive the effective static Hamiltonian 
Eq.~\eqref{eq:Hamil_eff} in the main text.
We apply the time-dependent unitary transform,
\begin{align}
 U\coloneqq e^{i\eta \Omega t S_{\mathrm{tot}}^{z}},
\end{align}
to $\mathcal{H}(t)$ as
\begin{align}
 \mathcal{H}(t)&\to 
   U[\mathcal{H}(t)-i\hbar\partial_{t}]U^{-1}
   \eqqcolon \mathcal{H}_{\mathrm{eff}}.
\end{align}
Then we obtain
the effective static Hamiltonian as
\begin{align}
 \mathcal{H}_{\mathrm{eff}}
   =\mathcal{H}_{0}-\eta\hbar\Omega S_{\mathrm{tot}}^{z}
   -B_{0}S_{\mathrm{tot}}^{x}.
\end{align}
In the case of weak laser field 
$B_{0}\ll \hbar\Omega$, 
the $B_{0}S_{\mathrm{tot}}^{x}$ term is negligibly small.
Thus we reach the effective static Hamiltonian 
Eq.~\eqref{eq:Hamil_eff} in the main text.

\section{Classical theory}
\label{sec:classical}

In this section, we derive the lower (upper) critical frequency 
$\Omega_{\mathrm{c1(c2)}}$. 
The classical spin configuration is determined 
in the way that the energy 
\begin{align}
 \epsilon
   =\frac{z_{0}J}{2}\boldsymbol{S}_{\mathrm{A}}\cdot\boldsymbol{S}_{\mathrm{B}}
     -\frac{D_{\mathrm{A}}}{2}(S_{\mathrm{A}}^{z})^{2}
     -\frac{D_{\mathrm{B}}}{2}(S_{\mathrm{B}}^{z})^{2}
     -\eta \frac{\hbar\Omega}{2}(S_{\mathrm{A}}^{z}+S_{\mathrm{B}}^{z})
\label{eq:Classical_E_SM}
\end{align}
takes minimum. 
Since Eq.~\eqref{eq:Classical_E_SM} has the $U(1)$ symmetry, 
we assume that $\boldsymbol{S}_{\mathrm{A}}$ and 
$\boldsymbol{S}_{\mathrm{B}}$ are in the $xz$ plane. 
We parametrize the spins as 
$\boldsymbol{S}_{\mathrm{A}}=( S_{\mathrm{A}}\sin\theta_{\mathrm{A}}, 0,
\eta S_{\mathrm{A}}\cos\theta_{\mathrm{A}})$ and 
$\boldsymbol{S}_{\mathrm{B}}=(-S_{\mathrm{B}}\sin\theta_{\mathrm{B}}, 0,
\eta S_{\mathrm{B}}\cos\theta_{\mathrm{B}})$.
Then Eq.~\eqref{eq:Classical_E_SM} can be rewritten as 
\begin{align}
 \epsilon
   =\frac{z_{0}JS_{\mathrm{A}}S_{\mathrm{B}}}{2}
     \cos(\theta_{\mathrm{A}}+\theta_{\mathrm{B}})
     -\frac{D_{\mathrm{A}}S_{\mathrm{A}}^{2}}{2}
       \cos^{2}\theta_{\mathrm{A}}
     -\frac{D_{\mathrm{B}}S_{\mathrm{B}}^{2}}{2}
       \cos^{2}\theta_{\mathrm{B}}
     -\frac{\hbar\Omega}{2}(S_{\mathrm{A}}\cos\theta_{\mathrm{A}}
       +S_{\mathrm{B}}\cos\theta_{\mathrm{B}}).
\end{align}
From the conditions for the energy minimum, 
$\partial\epsilon/\partial\theta_{\mathrm{A}}=0$ and 
$\partial\epsilon/\partial\theta_{\mathrm{B}}=0$, 
we obtain 
\begin{align}
 -z_{0}JS_{\mathrm{B}}
     \sin(\theta_{\mathrm{A}}+\theta_{\mathrm{B}})
     +2D_{\mathrm{A}}S_{\mathrm{A}}
       \cos\theta_{\mathrm{A}}\sin\theta_{\mathrm{A}}
     +\hbar\Omega\sin\theta_{\mathrm{A}}=&0,
\label{eq:MinCond1}\\
 -z_{0}JS_{\mathrm{A}}
     \sin(\theta_{\mathrm{A}}+\theta_{\mathrm{B}})
     +2D_{\mathrm{B}}S_{\mathrm{B}}
       \cos\theta_{\mathrm{B}}\sin\theta_{\mathrm{B}}
     +\hbar\Omega\sin\theta_{\mathrm{B}}=&0.
\label{eq:MinCond2}
\end{align}

\subsection{Around $\Omega=\Omega_{\mathrm{c}1}$}

We consider the frequency just above $\Omega_{\mathrm{c}1}$, 
where $\sin\theta_{\mathrm{A}}\simeq \theta_{\mathrm{A}}$, 
$\sin\theta_{\mathrm{B}}\simeq \pi-\theta_{\mathrm{B}}$, 
$\cos\theta_{\mathrm{A}}\simeq 1$, 
$\cos\theta_{\mathrm{B}}\simeq -1$, 
$\theta_{\mathrm{A}}\neq 0$, and $\pi-\theta_{\mathrm{B}}\neq 0$. 
Then Eqs.~\eqref{eq:MinCond1} and \eqref{eq:MinCond2} become 
\begin{align}
 &-z_{0}JS_{\mathrm{B}}
     (-\theta_{\mathrm{A}}+\pi-\theta_{\mathrm{B}})
     +2D_{\mathrm{A}}S_{\mathrm{A}}\theta_{\mathrm{A}}
     +\hbar\Omega_{\mathrm{c}1}\theta_{\mathrm{A}}=0
 \Leftrightarrow
 z_{0}JS_{\mathrm{B}}(\pi-\theta_{\mathrm{B}})/\theta_{\mathrm{A}}
   =z_{0}JS_{\mathrm{B}}
     +2D_{\mathrm{A}}S_{\mathrm{A}}
     +\hbar\Omega_{\mathrm{c}1},
\nonumber\\
 &-z_{0}JS_{\mathrm{A}}
     (-\theta_{\mathrm{A}}+\pi-\theta_{\mathrm{B}})
     -2D_{\mathrm{B}}S_{\mathrm{B}}(\pi-\theta_{\mathrm{B}})
     +\hbar\Omega_{\mathrm{c}1}(\pi-\theta_{\mathrm{B}})=0
 \Leftrightarrow
 z_{0}JS_{\mathrm{A}}\theta_{\mathrm{A}}/(\pi-\theta_{\mathrm{B}})
   =z_{0}JS_{\mathrm{A}}
     +2D_{\mathrm{B}}S_{\mathrm{B}}
     -\hbar\Omega_{\mathrm{c}1}.
\nonumber
\end{align}
Thus
\begin{align}
 &(\hbar\Omega_{\mathrm{c}1}+z_{0}JS_{\mathrm{B}}+2D_{\mathrm{A}}S_{\mathrm{A}})
   (\hbar\Omega_{\mathrm{c}1}-z_{0}JS_{\mathrm{A}}-2D_{\mathrm{B}}S_{\mathrm{B}})
     =-z_{0}^{2}J^{2}S_{\mathrm{A}}S_{\mathrm{B}}
\nonumber\\
 \Leftrightarrow
 &[\{\hbar\Omega_{\mathrm{c}1}-z_{0}J(S_{\mathrm{A}}-S_{\mathrm{B}})/2
   +D_{\mathrm{A}}S_{\mathrm{A}}-D_{\mathrm{B}}S_{\mathrm{B}}\}
   +\{z_{0}J(S_{\mathrm{A}}+S_{\mathrm{B}})/2
   +D_{\mathrm{A}}S_{\mathrm{A}}+D_{\mathrm{B}}S_{\mathrm{B}}\}]
\nonumber\\
   &\times
   [\{\hbar\Omega_{\mathrm{c}1}-z_{0}J(S_{\mathrm{A}}-S_{\mathrm{B}})/2
   +D_{\mathrm{A}}S_{\mathrm{A}}-D_{\mathrm{B}}S_{\mathrm{B}}\}
   -\{z_{0}J(S_{\mathrm{A}}+S_{\mathrm{B}})/2
   +D_{\mathrm{A}}S_{\mathrm{A}}+D_{\mathrm{B}}S_{\mathrm{B}}\}]
     =-z_{0}^{2}J^{2}S_{\mathrm{A}}S_{\mathrm{B}}
\nonumber\\
 \Leftrightarrow
 &\hbar\Omega_{\mathrm{c}1}
   =z_{0}J(S_{\mathrm{A}}-S_{\mathrm{B}})/2
   -D_{\mathrm{A}}S_{\mathrm{A}}+D_{\mathrm{B}}S_{\mathrm{B}}
   \pm\sqrt{-z_{0}^{2}J^{2}S_{\mathrm{A}}S_{\mathrm{B}}
     +[z_{0}J(S_{\mathrm{A}}+S_{\mathrm{B}})/2
     +D_{\mathrm{A}}S_{\mathrm{A}}+D_{\mathrm{B}}S_{\mathrm{B}}]^{2}}.
\nonumber
\end{align}
From $\Omega_{\mathrm{c}1}>0$, we obtain 
\begin{align}
 \hbar\Omega_{\mathrm{c}1}
   =z_{0}J(S_{\mathrm{A}}-S_{\mathrm{B}})/2
   -D_{\mathrm{A}}S_{\mathrm{A}}+D_{\mathrm{B}}S_{\mathrm{B}}
   +\sqrt{-z_{0}^{2}J^{2}S_{\mathrm{A}}S_{\mathrm{B}}
     +[z_{0}J(S_{\mathrm{A}}+S_{\mathrm{B}})/2
     +D_{\mathrm{A}}S_{\mathrm{A}}+D_{\mathrm{B}}S_{\mathrm{B}}]^{2}}.
\label{eq:Omegac1_SM}
\end{align}

\subsection{Around $\Omega=\Omega_{\mathrm{c}2}$}

We consider the frequency just below $\Omega_{\mathrm{c}2}$, 
where $\sin\theta_{\mathrm{A}}\simeq \theta_{\mathrm{A}}$, 
$\sin\theta_{\mathrm{B}}\simeq \theta_{\mathrm{B}}$, 
$\cos\theta_{\mathrm{A}}\simeq 1$, 
$\cos\theta_{\mathrm{B}}\simeq 1$, 
$\theta_{\mathrm{A}}\neq 0$, and $\theta_{\mathrm{B}}\neq 0$. 
Then Eqs.~\eqref{eq:MinCond1} and \eqref{eq:MinCond2} become 
\begin{align}
 &-z_{0}JS_{\mathrm{B}}
     (\theta_{\mathrm{A}}+\theta_{\mathrm{B}})
     +2D_{\mathrm{A}}S_{\mathrm{A}}\theta_{\mathrm{A}}
     +\hbar\Omega_{\mathrm{c}2}\theta_{\mathrm{A}}=0
 \Leftrightarrow
 z_{0}JS_{\mathrm{B}}\theta_{\mathrm{B}}/\theta_{\mathrm{A}}
   =-z_{0}JS_{\mathrm{B}}
     +2D_{\mathrm{A}}S_{\mathrm{A}}
     +\hbar\Omega_{\mathrm{c}2},
\nonumber\\
 &-z_{0}JS_{\mathrm{A}}
     (\theta_{\mathrm{A}}+\theta_{\mathrm{B}})
     +2D_{\mathrm{B}}S_{\mathrm{B}}\theta_{\mathrm{B}}
     +\hbar\Omega_{\mathrm{c}2}\theta_{\mathrm{B}}=0
 \Leftrightarrow
 z_{0}JS_{\mathrm{A}}\theta_{\mathrm{A}}/\theta_{\mathrm{B}}
   =-z_{0}JS_{\mathrm{A}}
     +2D_{\mathrm{B}}S_{\mathrm{B}}
     +\hbar\Omega_{\mathrm{c}2}.
\nonumber
\end{align}
Thus
\begin{align}
 &(\hbar\Omega_{\mathrm{c}2}-z_{0}JS_{\mathrm{B}}+2D_{\mathrm{A}}S_{\mathrm{A}})
   (\hbar\Omega_{\mathrm{c}2}-z_{0}JS_{\mathrm{A}}+2D_{\mathrm{B}}S_{\mathrm{B}})
     =z_{0}^{2}J^{2}S_{\mathrm{A}}S_{\mathrm{B}}
\nonumber\\
 \Leftrightarrow
 &[\{\hbar\Omega_{\mathrm{c}2}-z_{0}J(S_{\mathrm{A}}+S_{\mathrm{B}})/2
   +D_{\mathrm{A}}S_{\mathrm{A}}+D_{\mathrm{B}}S_{\mathrm{B}}\}
   +\{z_{0}J(S_{\mathrm{A}}-S_{\mathrm{B}})/2
   +D_{\mathrm{A}}S_{\mathrm{A}}-D_{\mathrm{B}}S_{\mathrm{B}}\}]
\nonumber\\
   &\times
   [\{\hbar\Omega_{\mathrm{c}2}-z_{0}J(S_{\mathrm{A}}+S_{\mathrm{B}})/2
   +D_{\mathrm{A}}S_{\mathrm{A}}+D_{\mathrm{B}}S_{\mathrm{B}}\}
   -\{z_{0}J(S_{\mathrm{A}}-S_{\mathrm{B}})/2
   +D_{\mathrm{A}}S_{\mathrm{A}}-D_{\mathrm{B}}S_{\mathrm{B}}\}]
     =z_{0}^{2}J^{2}S_{\mathrm{A}}S_{\mathrm{B}}
\nonumber\\
 \Leftrightarrow
 &\hbar\Omega_{\mathrm{c}2}
   =z_{0}J(S_{\mathrm{A}}+S_{\mathrm{B}})/2
   -D_{\mathrm{A}}S_{\mathrm{A}}-D_{\mathrm{B}}S_{\mathrm{B}}
   \pm\sqrt{z_{0}^{2}J^{2}S_{\mathrm{A}}S_{\mathrm{B}}
     +[z_{0}J(S_{\mathrm{A}}-S_{\mathrm{B}})/2
     +D_{\mathrm{A}}S_{\mathrm{A}}-D_{\mathrm{B}}S_{\mathrm{B}}]^{2}}.
\nonumber
\end{align}
From $\Omega_{\mathrm{c}2}>0$, we obtain
\begin{align}
 \hbar\Omega_{\mathrm{c}2}
   =z_{0}J(S_{\mathrm{A}}+S_{\mathrm{B}})/2
   -D_{\mathrm{A}}S_{\mathrm{A}}-D_{\mathrm{B}}S_{\mathrm{B}}
   +\sqrt{z_{0}^{2}J^{2}S_{\mathrm{A}}S_{\mathrm{B}}
     +[z_{0}J(S_{\mathrm{A}}-S_{\mathrm{B}})/2
     +D_{\mathrm{A}}S_{\mathrm{A}}-D_{\mathrm{B}}S_{\mathrm{B}}]^{2}}.
\label{eq:Omegac2_SM}
\end{align}

\section{Spin wave theory}
\label{sec:spinwave}

In this section, we derive the magnon BEC transition point 
$\Omega_{\mathrm{BEC1(BEC2)}}$ 
and see that it coincides with the lower (upper) critical frequency 
${\Omega}_{\mathrm{c1(c2)}}$. 
We consider the system 
\begin{align}
 {\mathcal{H}}_{\mathrm{eff}}
   =J\sum_{\langle i\in \mathrm{A}, j\in \mathrm{B} \rangle}
     \boldsymbol{S}_{\mathrm{A},i} \cdot
     \boldsymbol{S}_{\mathrm{B},j}
     -D_{\mathrm{A}}\sum_{i\in\mathrm{A}}
       (S_{\mathrm{A},i}^{z})^{2}
     -D_{\mathrm{B}}\sum_{j\in\mathrm{B}}
       (S_{\mathrm{B},j}^{z})^{2}
   -\eta\hbar\Omega\Big(
     \sum_{i\in\mathrm{A}}S_{\mathrm{A},i}^{z}
     +\sum_{j\in\mathrm{B}}S_{\mathrm{B},j}^{z}\Big).
\label{eq:Hamil_SM}
\end{align}
The boundary condition is periodic, and the number of sites is $N$;
$N/2$ sites for the A and B sublattice.

\subsection{Around $\Omega=\Omega_{\mathrm{BEC}1}$}
\label{subsec:omega1}

The ground state is ferrimagnetic 
$\boldsymbol{S}_{\mathrm{A}}=(0,0, \eta S_{\mathrm{A}})$ and 
$\boldsymbol{S}_{\mathrm{B}}=(0,0,-\eta S_{\mathrm{B}})$. 
We perform the Holstein-Primakoff transformation,
\begin{align}
 &\eta S_{\mathrm{A},i}^{z}
   =S_{\mathrm{A}}-n_{\mathrm{A},i},\quad
 S_{\mathrm{A},i}^{x}+\eta iS_{\mathrm{A},i}^{y}
   =\sqrt{2S_{\mathrm{A}}}
     \Big(1-\frac{n_{\mathrm{A},i}}{2S_{\mathrm{A}}}\Big)^{1/2}
     b_{\mathrm{A},i},\quad
 S_{\mathrm{A},i}^{x}-\eta iS_{\mathrm{A},i}^{y}
   =\sqrt{2S_{\mathrm{A}}}b_{\mathrm{A},i}^{\dagger}
     \Big(1-\frac{n_{\mathrm{A},i}}{2S_{\mathrm{A}}}\Big)^{1/2},
\nonumber\\
 &\eta S_{\mathrm{B},j}^{z}
   =-S_{\mathrm{B}}+n_{\mathrm{B},j},\quad
 S_{\mathrm{B},j}^{x}+\eta iS_{\mathrm{B},j}^{y}
   =\sqrt{2S_{\mathrm{B}}}b_{\mathrm{B},j}^{\dagger}
     \Big(1-\frac{n_{\mathrm{B},j}}{2S_{\mathrm{B}}}\Big)^{1/2},
 S_{\mathrm{B},j}^{x}-\eta iS_{\mathrm{B},j}^{y}
   =\sqrt{2S_{\mathrm{B}}}
     \Big(1-\frac{n_{\mathrm{B},j}}{2S_{\mathrm{B}}}\Big)^{1/2}
     b_{\mathrm{B},j},\quad
\nonumber
\end{align}
where $b_{j}^{\dagger}$ and $b_{j}$ are 
creation and annihilation operators for bosons (magnons), 
and $n_{(\mathrm{A},\mathrm{B}),j}\equiv
b_{(\mathrm{A},\mathrm{B}),j}^{\dagger}b_{(\mathrm{A},\mathrm{B}),j}$ 
is the number operator. 
We make an expansion and 
retain up to the second order in terms of $b$ and $b^{\dagger}$, 
\begin{align}
 &\eta S_{\mathrm{A},i}^{z}
   =S_{\mathrm{A}}-n_{\mathrm{A},i},\quad
 S_{\mathrm{A},i}^{x}+\eta iS_{\mathrm{A},i}^{y}
   =\sqrt{2S_{\mathrm{A}}}
     b_{\mathrm{A},i},\quad
 S_{\mathrm{A},i}^{x}-\eta iS_{\mathrm{A},i}^{y}
   =\sqrt{2S_{\mathrm{A}}}b_{\mathrm{A},i}^{\dagger},
\nonumber\\
 &\eta S_{\mathrm{B},j}^{z}
   =-S_{\mathrm{B}}+n_{\mathrm{B},j},\quad
 S_{\mathrm{B},j}^{x}+\eta iS_{\mathrm{B},j}^{y}
   =\sqrt{2S_{\mathrm{B}}}
     b_{\mathrm{B},j}^{\dagger},\quad
 S_{\mathrm{B},j}^{x}-\eta iS_{\mathrm{B},j}^{y}
   =\sqrt{2S_{\mathrm{B}}}b_{\mathrm{B},j}.
\nonumber
\end{align}
Using magnon operators, 
the Hamiltonian~\eqref{eq:Hamil_SM} is rewritten as 
\begin{align}
 {\mathcal{H}}_{\mathrm{eff}}
   =&J\sqrt{S_{\mathrm{A}}S_{\mathrm{B}}}
     \sum_{\langle i\in \mathrm{A}, j\in \mathrm{B} \rangle}
     (b_{\mathrm{A},i}b_{\mathrm{B},j}+\mathrm{H.c.})
   +z_{0}JS_{\mathrm{B}}\sum_{i\in\mathrm{A}}n_{\mathrm{A},i}
   +z_{0}JS_{\mathrm{A}}\sum_{j\in\mathrm{B}}n_{\mathrm{B},j}
\nonumber\\
   &+2D_{\mathrm{A}}S_{\mathrm{A}}
     \sum_{i\in\mathrm{A}}n_{\mathrm{A},i}
   +2D_{\mathrm{B}}S_{\mathrm{B}}
     \sum_{j\in\mathrm{B}}n_{\mathrm{B},j}
   +\hbar\Omega\Big(
     \sum_{i\in\mathrm{A}}n_{\mathrm{A},i}
     -\sum_{j\in\mathrm{B}}n_{\mathrm{B},j}\Big),
\end{align}
where the constant terms are dropped. 
We consider the cubic lattice and the coordination number is $z_{0}=6$. 
After the Fourier transform 
\begin{align}
 &b_{\mathrm{A},\boldsymbol{k}}
   =\sqrt{\frac{2}{N}}\sum_{i\in\mathrm{A}}
     e^{-i\boldsymbol{k}\cdot\boldsymbol{r}_{i}}
     b_{\mathrm{A},i},\quad
 b_{\mathrm{A},\boldsymbol{k}}^{\dagger}
   =\sqrt{\frac{2}{N}}\sum_{i\in\mathrm{A}}
     e^{i\boldsymbol{k}\cdot\boldsymbol{r}_{i}}
     b_{\mathrm{A},i}^{\dagger},\quad
 n_{\mathrm{A},\boldsymbol{k}}
   =b_{\mathrm{A},\boldsymbol{k}}^{\dagger}
     b_{\mathrm{A},\boldsymbol{k}},\nonumber\\
 &b_{\mathrm{B},\boldsymbol{k}}
   =\sqrt{\frac{2}{N}}\sum_{i\in\mathrm{B}}
     e^{i\boldsymbol{k}\cdot\boldsymbol{r}_{i}}
     b_{\mathrm{B},i},\quad
 b_{\mathrm{B},\boldsymbol{k}}^{\dagger}
   =\sqrt{\frac{2}{N}}\sum_{i\in\mathrm{B}}
     e^{-i\boldsymbol{k}\cdot\boldsymbol{r}_{i}}
     b_{\mathrm{B},i}^{\dagger},\quad
 n_{\mathrm{B},\boldsymbol{k}}
   =b_{\mathrm{B},\boldsymbol{k}}^{\dagger}
     b_{\mathrm{B},\boldsymbol{k}},
\nonumber
\end{align}
($\boldsymbol{r}_{i}$ is the positional vector), 
we obtain 
\begin{align}
 {\mathcal{H}}_{\mathrm{eff}}
   =&J\sqrt{S_{\mathrm{A}}S_{\mathrm{B}}}
     \sum_{\boldsymbol{k}}
     2[\cos(k_{x}a_{0})+\cos(k_{y}a_{0})+\cos(k_{z}a_{0})]
     (b_{\mathrm{A},\boldsymbol{k}}b_{\mathrm{B},\boldsymbol{k}}
       +\mathrm{H.c.})
\nonumber\\
   &+(z_{0}JS_{\mathrm{B}}+2D_{\mathrm{A}}S_{\mathrm{A}}+\hbar\Omega)
     \sum_{\boldsymbol{k}}n_{\mathrm{A},\boldsymbol{k}}
   +(z_{0}JS_{\mathrm{A}}+2D_{\mathrm{B}}S_{\mathrm{B}}-\hbar\Omega)
     \sum_{\boldsymbol{k}}n_{\mathrm{B},\boldsymbol{k}},
\nonumber
\end{align}
where $a_{0}$ is the lattice constant. 
We perform the Bogoliubov transformation 
\begin{align}
\begin{pmatrix}
 \alpha_{+,\boldsymbol{k}} \\
 \alpha_{-,\boldsymbol{k}}^{\dagger}
\end{pmatrix}
=
\begin{pmatrix}
 \cosh\theta_{\boldsymbol{k}} & \sinh\theta_{\boldsymbol{k}} \\
 \sinh\theta_{\boldsymbol{k}} & \cosh\theta_{\boldsymbol{k}}
\end{pmatrix}
\begin{pmatrix}
 b_{\mathrm{A},\boldsymbol{k}} \\
 b_{\mathrm{B},\boldsymbol{k}}^{\dagger}
\end{pmatrix}
,\nonumber
\end{align}
with the angle 
\begin{align}
 \tanh 2\theta_{\boldsymbol{k}}
   =\frac{2f(\boldsymbol{k})}{C_{1}+C_{2}},\nonumber
\end{align}
where
\begin{align}
 f(\boldsymbol{k})=&2J\sqrt{S_{\mathrm{A}}S_{\mathrm{B}}}
   [\cos(k_{x}a_{0})+\cos(k_{y}a_{0})+\cos(k_{z}a_{0})],
\nonumber\\
 C_{1}=&z_{0}JS_{\mathrm{B}}+2D_{\mathrm{A}}S_{\mathrm{A}}+\hbar\Omega,
\nonumber\\
 C_{2}=&z_{0}JS_{\mathrm{A}}+2D_{\mathrm{B}}S_{\mathrm{B}}-\hbar\Omega.
\nonumber
\end{align}
Then the Hamiltonian becomes 
\begin{align}
 {\mathcal{H}}_{\mathrm{eff}}
   =&\sum_{k}
     \Big(-f(\boldsymbol{k})\sinh 2\theta_{\boldsymbol{k}}
     +\frac{C_{1}-C_{2}}{2}
     +\frac{C_{1}+C_{2}}{2}\cosh 2\theta_{\boldsymbol{k}}\Big)
     \alpha_{+,\boldsymbol{k}}^{\dagger}\alpha_{+,\boldsymbol{k}}
\nonumber\\
   &+\sum_{k}
     \Big(-f(\boldsymbol{k})\sinh 2\theta_{\boldsymbol{k}}
     -\frac{C_{1}-C_{2}}{2}
     +\frac{C_{1}+C_{2}}{2}\cosh 2\theta_{\boldsymbol{k}}\Big)
     \alpha_{-,\boldsymbol{k}}^{\dagger}\alpha_{-,\boldsymbol{k}}
\nonumber\\
   =&\sum_{k}
     \Big[\frac{C_{1}-C_{2}}{2}
     +\sqrt{-f(\boldsymbol{k})^{2}+\Big(\frac{C_{1}+C_{2}}{2}\Big)^{2}}\Big]
     \alpha_{+,\boldsymbol{k}}^{\dagger}\alpha_{+,\boldsymbol{k}}
\nonumber\\
   &+\sum_{k}
     \Big[-\frac{C_{1}-C_{2}}{2}
     +\sqrt{-f(\boldsymbol{k})^{2}+\Big(\frac{C_{1}+C_{2}}{2}\Big)^{2}}\Big]
     \alpha_{-,\boldsymbol{k}}^{\dagger}\alpha_{-,\boldsymbol{k}},
\end{align}
where the constant terms are dropped. 
We can rewrite the Hamiltonian in the form 
\begin{align}
 {\mathcal{H}}_{\mathrm{eff}}=\sum_{\sigma=\pm,\boldsymbol{k}}
   (\hbar\omega_{\sigma,\boldsymbol{k}}^{[\alpha]}
     +\Delta_{\sigma}^{[\alpha]}
     +\sigma\hbar\Omega)
   \alpha_{\sigma,\boldsymbol{k}}^{\dagger}\alpha_{\sigma,\boldsymbol{k}},
\label{eq:Hamil_MagBEC1_SM}
\end{align}
where $\hbar\omega_{\sigma,\boldsymbol{k}}^{[\alpha]}$ is the energy dispersion
and $\Delta_{\sigma}^{[\alpha]}+\sigma\hbar\Omega$ is the magnon gap in laser
represented as 
\begin{align}
 \hbar\omega_{\pm,\boldsymbol{k}}^{[\alpha]}
   =&\sqrt{-f(\boldsymbol{k})^{2}
     +[3J(S_{\mathrm{A}}+S_{\mathrm{B}})
     +D_{\mathrm{A}}S_{\mathrm{A}}+D_{\mathrm{B}}S_{\mathrm{B}}]^{2}}
   -\sqrt{-f(\boldsymbol{0})^{2}
     +[3J(S_{\mathrm{A}}+S_{\mathrm{B}})
     +D_{\mathrm{A}}S_{\mathrm{A}}+D_{\mathrm{B}}S_{\mathrm{B}}]^{2}},
\nonumber\\
 \Delta_{\pm}^{[\alpha]}
   =&\mp[3J(S_{\mathrm{A}}-S_{\mathrm{B}})
     -D_{\mathrm{A}}S_{\mathrm{A}}+D_{\mathrm{B}}S_{\mathrm{B}}]
     +\sqrt{-f(\boldsymbol{0})^{2}
     +[3J(S_{\mathrm{A}}+S_{\mathrm{B}})
     +D_{\mathrm{A}}S_{\mathrm{A}}+D_{\mathrm{B}}S_{\mathrm{B}}]^{2}},
\nonumber
\end{align}
noting that $f(\boldsymbol{k})$ takes the maximum at 
$\boldsymbol{k}=\boldsymbol{0}$. 
Therefore, when $\Omega$ is increased from the small value, 
the magnon created by 
$\alpha_{-,\boldsymbol{k}=\boldsymbol{0}}^{\dagger}$ condensates at 
\begin{align}
 \hbar\Omega_{\mathrm{BEC1}}=\Delta_{-}^{[\alpha]}
   =3J(S_{\mathrm{A}}-S_{\mathrm{B}})
   -D_{\mathrm{A}}S_{\mathrm{A}}+D_{\mathrm{B}}S_{\mathrm{B}}
   +\sqrt{-36J^{2}S_{\mathrm{A}}S_{\mathrm{B}}
     +[3J(S_{\mathrm{A}}+S_{\mathrm{B}})
     +D_{\mathrm{A}}S_{\mathrm{A}}+D_{\mathrm{B}}S_{\mathrm{B}}]^{2}},
\end{align}
which agrees with $\hbar\Omega_{\mathrm{c1}}$ 
[Eq.~\eqref{eq:Omegac1_SM}].

\subsection{Around $\Omega=\Omega_{\mathrm{BEC}2}$}
\label{subsec:omega2}

The ground state is ferromagnetic 
$\boldsymbol{S}_{\mathrm{A}}=(0,0,\eta S_{\mathrm{A}})$ and 
$\boldsymbol{S}_{\mathrm{B}}=(0,0,\eta S_{\mathrm{B}})$. 
We perform the Holstein-Primakoff transformation,
\begin{align}
 &\eta S_{\mathrm{A},i}^{z}
   =S_{\mathrm{A}}-n_{\mathrm{A},i},\quad
 S_{\mathrm{A},i}^{x}+\eta iS_{\mathrm{A},i}^{y}
   =\sqrt{2S_{\mathrm{A}}}
     \Big(1-\frac{n_{\mathrm{A},i}}{2S_{\mathrm{A}}}\Big)^{1/2}
     b_{\mathrm{A},i},\quad
 S_{\mathrm{A},i}^{x}-\eta iS_{\mathrm{A},i}^{y}
   =\sqrt{2S_{\mathrm{A}}}b_{\mathrm{A},i}^{\dagger}
     \Big(1-\frac{n_{\mathrm{A},i}}{2S_{\mathrm{A}}}\Big)^{1/2},
\nonumber\\
 &\eta S_{\mathrm{B},j}^{z}
   =S_{\mathrm{B}}-n_{\mathrm{B},j},\quad
 S_{\mathrm{B},j}^{x}+\eta iS_{\mathrm{B},j}^{y}
   =\sqrt{2S_{\mathrm{B}}}
     \Big(1-\frac{n_{\mathrm{B},j}}{2S_{\mathrm{B}}}\Big)^{1/2}
     b_{\mathrm{B},j},\quad
 S_{\mathrm{B},j}^{x}-\eta iS_{\mathrm{B},j}^{y}
   =\sqrt{2S_{\mathrm{B}}}b_{\mathrm{B},j}^{\dagger}
     \Big(1-\frac{n_{\mathrm{B},j}}{2S_{\mathrm{B}}}\Big)^{1/2},
\nonumber
\end{align}
where $b_{j}^{\dagger}$ and $b_{j}$ are 
creation and annihilation operators for bosons (magnons), 
and $n_{(\mathrm{A},\mathrm{B}),j}\equiv
b_{(\mathrm{A},\mathrm{B}),j}^{\dagger}b_{(\mathrm{A},\mathrm{B}),j}$ 
is the number operator. 
We make an expansion and 
retain up to the second order in terms of $b$ and $b^{\dagger}$, 
\begin{align}
 &\eta S_{\mathrm{A},i}^{z}
   =S_{\mathrm{A}}-n_{\mathrm{A},i},\quad
 S_{\mathrm{A},i}^{x}+\eta iS_{\mathrm{A},i}^{y}
   =\sqrt{2S_{\mathrm{A}}}
     b_{\mathrm{A},i},\quad
 S_{\mathrm{A},i}^{x}-\eta iS_{\mathrm{A},i}^{y}
   =\sqrt{2S_{\mathrm{A}}}b_{\mathrm{A},i}^{\dagger},
\nonumber\\
 &\eta S_{\mathrm{B},j}^{z}
   =S_{\mathrm{B}}-n_{\mathrm{B},j},\quad
 S_{\mathrm{B},j}^{x}+\eta iS_{\mathrm{B},j}^{y}
   =\sqrt{2S_{\mathrm{B}}}
     b_{\mathrm{B},j},\quad
 S_{\mathrm{B},j}^{x}-\eta iS_{\mathrm{B},j}^{y}
   =\sqrt{2S_{\mathrm{B}}}b_{\mathrm{B},j}^{\dagger}.
\nonumber
\end{align}
Using magnon operators, 
the Hamiltonian~\eqref{eq:Hamil_SM} is rewritten as 
\begin{align}
 {\mathcal{H}}_{\mathrm{eff}}
   =&J\sqrt{S_{\mathrm{A}}S_{\mathrm{B}}}
     \sum_{\langle i\in \mathrm{A}, j\in \mathrm{B} \rangle}
     (b_{\mathrm{A},i}^{\dagger}b_{\mathrm{B},j}+\mathrm{H.c.})
   -z_{0}JS_{\mathrm{B}}\sum_{i\in\mathrm{A}}n_{\mathrm{A},i}
   -z_{0}JS_{\mathrm{A}}\sum_{j\in\mathrm{B}}n_{\mathrm{B},j}
\nonumber\\
   &+2D_{\mathrm{A}}S_{\mathrm{A}}
     \sum_{i\in\mathrm{A}}n_{\mathrm{A},i}
   +2D_{\mathrm{B}}S_{\mathrm{B}}
     \sum_{j\in\mathrm{B}}n_{\mathrm{B},j}
   +\hbar\Omega\Big(
     \sum_{i\in\mathrm{A}}n_{\mathrm{A},i}
     +\sum_{j\in\mathrm{B}}n_{\mathrm{B},j}\Big),
\end{align}
where the constant terms are dropped. 
We consider the cubic lattice and the coordination number is $z_{0}=6$. 
After the Fourier transform 
\begin{align}
 &b_{\mathrm{A},\boldsymbol{k}}
   =\sqrt{\frac{2}{N}}\sum_{i\in\mathrm{A}}
     e^{-i\boldsymbol{k}\cdot\boldsymbol{r}_{i}}
     b_{\mathrm{A},i},\quad
 b_{\mathrm{A},\boldsymbol{k}}^{\dagger}
   =\sqrt{\frac{2}{N}}\sum_{i\in\mathrm{A}}
     e^{i\boldsymbol{k}\cdot\boldsymbol{r}_{i}}
     b_{\mathrm{A},i}^{\dagger},\quad
 n_{\mathrm{A},\boldsymbol{k}}
   =b_{\mathrm{A},\boldsymbol{k}}^{\dagger}
     b_{\mathrm{A},\boldsymbol{k}},\nonumber\\
 &b_{\mathrm{B},\boldsymbol{k}}
   =\sqrt{\frac{2}{N}}\sum_{i\in\mathrm{B}}
     e^{-i\boldsymbol{k}\cdot\boldsymbol{r}_{i}}
     b_{\mathrm{B},i},\quad
 b_{\mathrm{B},\boldsymbol{k}}^{\dagger}
   =\sqrt{\frac{2}{N}}\sum_{i\in\mathrm{B}}
     e^{i\boldsymbol{k}\cdot\boldsymbol{r}_{i}}
     b_{\mathrm{B},i}^{\dagger},\quad
 n_{\mathrm{B},\boldsymbol{k}}
   =b_{\mathrm{B},\boldsymbol{k}}^{\dagger}
     b_{\mathrm{B},\boldsymbol{k}},
\nonumber
\end{align}
($\boldsymbol{r}_{i}$ is the positional vector), 
we obtain 
\begin{align}
 {\mathcal{H}}_{\mathrm{eff}}
   =&J\sqrt{S_{\mathrm{A}}S_{\mathrm{B}}}
     \sum_{\boldsymbol{k}}
     2[\cos(k_{x}a_{0})+\cos(k_{y}a_{0})+\cos(k_{z}a_{0})]
     (b_{\mathrm{A},\boldsymbol{k}}^{\dagger}b_{\mathrm{B},\boldsymbol{k}}
       +\mathrm{H.c.})
\nonumber\\
   &+(-z_{0}JS_{\mathrm{B}}+2D_{\mathrm{A}}S_{\mathrm{A}}+\hbar\Omega)
     \sum_{\boldsymbol{k}}n_{\mathrm{A},\boldsymbol{k}}
   +(-z_{0}JS_{\mathrm{A}}+2D_{\mathrm{B}}S_{\mathrm{B}}+\hbar\Omega)
     \sum_{\boldsymbol{k}}n_{\mathrm{B},\boldsymbol{k}},
\nonumber
\end{align}
where $a_{0}$ is the lattice constant. 
We perform the transformation 
\begin{align}
\begin{pmatrix}
 \beta_{+,\boldsymbol{k}} \\
 \beta_{-,\boldsymbol{k}}
\end{pmatrix}
=
\begin{pmatrix}
 \cos\theta_{\boldsymbol{k}} & -\sin\theta_{\boldsymbol{k}} \\
 \sin\theta_{\boldsymbol{k}} &  \cos\theta_{\boldsymbol{k}}
\end{pmatrix}
\begin{pmatrix}
 b_{\mathrm{A},\boldsymbol{k}} \\
 b_{\mathrm{B},\boldsymbol{k}}
\end{pmatrix}
,\nonumber
\end{align}
with the angle 
\begin{align}
 \tan 2\theta_{\boldsymbol{k}}
   =-\frac{2f(\boldsymbol{k})}{C_{1}-C_{2}},\nonumber
\end{align}
where
\begin{align}
 f(\boldsymbol{k})=&2J\sqrt{S_{\mathrm{A}}S_{\mathrm{B}}}
   [\cos(k_{x}a_{0})+\cos(k_{y}a_{0})+\cos(k_{z}a_{0})],
\nonumber\\
 C_{1}=&-z_{0}JS_{\mathrm{B}}+2D_{\mathrm{A}}S_{\mathrm{A}}+\hbar\Omega,
\nonumber\\
 C_{2}=&-z_{0}JS_{\mathrm{A}}+2D_{\mathrm{B}}S_{\mathrm{B}}+\hbar\Omega.
\nonumber
\end{align}
Then the Hamiltonian becomes 
\begin{align}
 {\mathcal{H}}_{\mathrm{eff}}
   =&\sum_{k}
     \Big(-f(\boldsymbol{k})\sin 2\theta_{\boldsymbol{k}}
     +\frac{C_{1}+C_{2}}{2}
     +\frac{C_{1}-C_{2}}{2}\cos 2\theta_{\boldsymbol{k}}\Big)
     \beta_{+,\boldsymbol{k}}^{\dagger}\beta_{+,\boldsymbol{k}}
\nonumber\\
   &+\sum_{k}
     \Big(f(\boldsymbol{k})\sin 2\theta_{\boldsymbol{k}}
     +\frac{C_{1}+C_{2}}{2}
     -\frac{C_{1}-C_{2}}{2}\cos 2\theta_{\boldsymbol{k}}\Big)
     \beta_{-,\boldsymbol{k}}^{\dagger}\beta_{-,\boldsymbol{k}}
\nonumber\\
   =&\sum_{k}
     \Big[\frac{C_{1}+C_{2}}{2}
     +\sqrt{f(\boldsymbol{k})^{2}+\Big(\frac{C_{1}-C_{2}}{2}\Big)^{2}}\Big]
     \beta_{+,\boldsymbol{k}}^{\dagger}\beta_{+,\boldsymbol{k}}
\nonumber\\
   &+\sum_{k}
     \Big[\frac{C_{1}+C_{2}}{2}
     -\sqrt{f(\boldsymbol{k})^{2}+\Big(\frac{C_{1}-C_{2}}{2}\Big)^{2}}\Big]
     \beta_{-,\boldsymbol{k}}^{\dagger}\beta_{-,\boldsymbol{k}}.
\end{align}
We can rewrite the Hamiltonian in the form 
\begin{align}
 {\mathcal{H}}_{\mathrm{eff}}=\sum_{\sigma=\pm,\boldsymbol{k}}
   (\hbar\omega_{\sigma,\boldsymbol{k}}^{[\beta]}
     +\Delta_{\sigma}^{[\beta]}
     +\hbar\Omega)
   \beta_{\sigma,\boldsymbol{k}}^{\dagger}\beta_{\sigma,\boldsymbol{k}},
\label{eq:Hamil_MagBEC2_SM}
\end{align}
where $\hbar\omega_{\sigma,\boldsymbol{k}}^{[\beta]}$ is the energy dispersion
and $\Delta_{\sigma}^{[\beta]}+\hbar\Omega$ is the magnon gap in laser represented as 
\begin{align}
 \hbar\omega_{+,\boldsymbol{k}}^{[\beta]}
   =&\sqrt{f(\boldsymbol{k})^{2}
     +[3J(S_{\mathrm{A}}-S_{\mathrm{B}})
     +D_{\mathrm{A}}S_{\mathrm{A}}-D_{\mathrm{B}}S_{\mathrm{B}}]^{2}}
   -|3J(S_{\mathrm{A}}-S_{\mathrm{B}})
     +D_{\mathrm{A}}S_{\mathrm{A}}-D_{\mathrm{B}}S_{\mathrm{B}}|,
\nonumber\\
 \hbar\omega_{-,\boldsymbol{k}}^{[\beta]}
   =&\sqrt{f(\boldsymbol{0})^{2}
     +[3J(S_{\mathrm{A}}-S_{\mathrm{B}})
     +D_{\mathrm{A}}S_{\mathrm{A}}-D_{\mathrm{B}}S_{\mathrm{B}}]^{2}}
   -\sqrt{f(\boldsymbol{k})^{2}
     +[3J(S_{\mathrm{A}}-S_{\mathrm{B}})
     +D_{\mathrm{A}}S_{\mathrm{A}}-D_{\mathrm{B}}S_{\mathrm{B}}]^{2}},
\nonumber\\
 \Delta_{+}^{[\beta]}
   =&-3J(S_{\mathrm{A}}+S_{\mathrm{B}})
     +D_{\mathrm{A}}S_{\mathrm{A}}+D_{\mathrm{B}}S_{\mathrm{B}}
     +|3J(S_{\mathrm{A}}-S_{\mathrm{B}})
     +D_{\mathrm{A}}S_{\mathrm{A}}-D_{\mathrm{B}}S_{\mathrm{B}}|,
\nonumber\\
 \Delta_{-}^{[\beta]}
   =&-3J(S_{\mathrm{A}}+S_{\mathrm{B}})
     +D_{\mathrm{A}}S_{\mathrm{A}}+D_{\mathrm{B}}S_{\mathrm{B}}
     -\sqrt{f(\boldsymbol{0})^{2}
     +[3J(S_{\mathrm{A}}-S_{\mathrm{B}})
     +D_{\mathrm{A}}S_{\mathrm{A}}-D_{\mathrm{B}}S_{\mathrm{B}}]^{2}},
\nonumber
\end{align}
noting that $f(\boldsymbol{k})$ takes the maximum at 
$\boldsymbol{k}=\boldsymbol{0}$. 
Therefore, when $\Omega$ is decreased from the large value, 
the magnons created by 
$\beta_{-,\boldsymbol{k}=0}^{\dagger}$ condensate at
\begin{align}
 \hbar\Omega_{\mathrm{BEC2}}= -\Delta_{-}^{[\beta]}
   =3J(S_{\mathrm{A}}+S_{\mathrm{B}})
   -D_{\mathrm{A}}S_{\mathrm{A}}-D_{\mathrm{B}}S_{\mathrm{B}}
   +\sqrt{36J^{2}S_{\mathrm{A}}S_{\mathrm{B}}
     +[3J(S_{\mathrm{A}}-S_{\mathrm{B}})
     +D_{\mathrm{A}}S_{\mathrm{A}}-D_{\mathrm{B}}S_{\mathrm{B}}]^{2}},
\end{align}
which agrees with $\hbar\Omega_{\mathrm{c2}}$ 
[Eq.~\eqref{eq:Omegac2_SM}]. 
Note that $\Delta_{\sigma}^{[\beta]}$ takes the negative value $\Delta_{\sigma}^{[\beta]} \leq 0$.

We remark that 
in the case of an insulating \textit{ferromagnet}, 
the application of the circularly polarized laser increases the magnon gap 
and the optical magnon BEC does not occur.

\section{Time-dependent mean field theory}
\label{sec:tdMFT}

In this section,
we discuss the time evolution of sublattice magnetization.
To this end, we numerically simulate the dynamics of the system 
using the time-dependent mean field theory and 
recasting the equation of motion into the form 
\begin{align}
 \frac{d\boldsymbol{m}_{\mathrm{A}}}{dt}
   =\boldsymbol{m}_{\mathrm{A}}\times
     \boldsymbol{H}_{\mathrm{A}}^{\mathrm{MF}},\quad
 \frac{d\boldsymbol{m}_{\mathrm{B}}}{dt}
   =\boldsymbol{m}_{\mathrm{B}}\times
     \boldsymbol{H}_{\mathrm{B}}^{\mathrm{MF}}.
\label{eq:EoM_SM}
\end{align}
We treat $\boldsymbol{m}_{\mathrm{A}}$ and 
$\boldsymbol{m}_{\mathrm{B}}$ as classical vectors, 
then Eq.~\eqref{eq:EoM_SM} is nothing but 
the two-body Landau-Lifshitz-Gilbert equation. 
Here we assume the laser-induced phenomena is much faster 
than magnetization damping, and neglect the Gilbert term. 
From the time-dependent Hamiltonian, 
\begin{align}
 {\mathcal{H}}(t)
   =J\sum_{\langle i\in \mathrm{A}, j\in \mathrm{B} \rangle}
     \boldsymbol{S}_{\mathrm{A},i} \cdot
     \boldsymbol{S}_{\mathrm{B},j}
     -D_{\mathrm{A}}\sum_{i\in\mathrm{A}}
       (S_{\mathrm{A},i}^{z})^{2}
     -D_{\mathrm{B}}\sum_{j\in\mathrm{B}}
       (S_{\mathrm{B},j}^{z})^{2}
   -B_{0}[S_{\mathrm{tot}}^{x}\cos(\Omega t)
     +\eta S_{\mathrm{tot}}^{y}\sin(\Omega t)],
\end{align}
we can derive the mean fields as 
\begin{align}
 \boldsymbol{H}_{\mathrm{A}}^{\mathrm{MF}}=
\begin{pmatrix}
 -z_{0}J m_{\mathrm{B}}^{x}+B_{0}\cos(\Omega t)\\
 -z_{0}J m_{\mathrm{B}}^{y}+B_{0}\sin(\Omega t)\\
 -z_{0}J m_{\mathrm{B}}^{z}+2D_{\mathrm{A}}m_{\mathrm{A}}^{z}
\end{pmatrix}
 ,\quad\boldsymbol{H}_{\mathrm{B}}^{\mathrm{MF}}=
\begin{pmatrix}
 -z_{0}J m_{\mathrm{A}}^{x}+B_{0}\cos(\Omega t)\\
 -z_{0}J m_{\mathrm{A}}^{y}+B_{0}\sin(\Omega t)\\
 -z_{0}J m_{\mathrm{A}}^{z}+2D_{\mathrm{B}}m_{\mathrm{B}}^{z}
\end{pmatrix}
.\nonumber
\end{align}

\section{The optical Barnett field without chirping}
\label{sec:highfreq}

In this section,
we discuss the laser application without chirping.
In order to study the application of circularly polarized laser 
without chirping, the framework of the Floquet theory and 
the inverse frequency expansion can be utilized. 
This method is applicable for the high frequency region. 
The total Hamiltonian 
\begin{align}
 \mathcal{H}(t)=\mathcal{H}_{0}
   -\frac{B_{0}}{2}(e^{-i\Omega t}S_{\mathrm{tot}}^{\eta}
      +e^{i\Omega t}S_{\mathrm{tot}}^{-\eta}),
\end{align}
is temporally periodic and can be written in the form of 
\begin{align}
 \mathcal{H}(t)=\sum_{m\in\mathbf{Z}}
   H_{m}e^{im\Omega t},
\end{align}
where 
\begin{align}
 H_{0}=\mathcal{H}_{0},\quad
 H_{\pm 1}=-\frac{B_{0}}{2}S_{\mathrm{tot}}^{\mp\eta},\quad
 H_{|m|\geq 2}=0.\nonumber
\end{align}
In the inverse frequency expansion up to the $1/\Omega$ order, 
the Floquet effective Hamiltonian in the high frequency regime 
is provided as 
\begin{subequations}
\begin{align}
 \mathcal{H}_{\mathrm{HF}}
   =&H_{0}+\frac{1}{\hbar\Omega}
     \sum_{m=1}^{\infty}\frac{[H_{m},H_{-m}]}{m}
     +O(\Omega^{-2}), \\
   =&\mathcal{H}_{0}
     -\frac{\eta B_{0}^{2}}{2\hbar\Omega}S_{\mathrm{tot}}^{z}
     +O(\Omega^{-2}).
\end{align}
\end{subequations}
Thus the optical Barnett field, 
\begin{align}
 \mathcal{B}_{\mathrm{HF}}
   =\frac{B_{0}^{2}}{2\hbar^{2}\gamma\Omega},
\end{align}
is proportional to $1/\Omega$ and $B_{0}^{2}$. 
This analysis indicates that although the induced field is small, 
the optical Barnett effect still occurs in the high frequency region 
away from the adiabatic regime considered in the main text.

\end{widetext}


\begin{thebibliography}{66}%
\makeatletter
\providecommand \@ifxundefined [1]{%
 \@ifx{#1\undefined}
}%
\providecommand \@ifnum [1]{%
 \ifnum #1\expandafter \@firstoftwo
 \else \expandafter \@secondoftwo
 \fi
}%
\providecommand \@ifx [1]{%
 \ifx #1\expandafter \@firstoftwo
 \else \expandafter \@secondoftwo
 \fi
}%
\providecommand \natexlab [1]{#1}%
\providecommand \enquote  [1]{``#1''}%
\providecommand \bibnamefont  [1]{#1}%
\providecommand \bibfnamefont [1]{#1}%
\providecommand \citenamefont [1]{#1}%
\providecommand \href@noop [0]{\@secondoftwo}%
\providecommand \href [0]{\begingroup \@sanitize@url \@href}%
\providecommand \@href[1]{\@@startlink{#1}\@@href}%
\providecommand \@@href[1]{\endgroup#1\@@endlink}%
\providecommand \@sanitize@url [0]{\catcode `\\12\catcode `\$12\catcode
  `\&12\catcode `\#12\catcode `\^12\catcode `\_12\catcode `\%12\relax}%
\providecommand \@@startlink[1]{}%
\providecommand \@@endlink[0]{}%
\providecommand \url  [0]{\begingroup\@sanitize@url \@url }%
\providecommand \@url [1]{\endgroup\@href {#1}{\urlprefix }}%
\providecommand \urlprefix  [0]{URL }%
\providecommand \Eprint [0]{\href }%
\providecommand \doibase [0]{http://dx.doi.org/}%
\providecommand \selectlanguage [0]{\@gobble}%
\providecommand \bibinfo  [0]{\@secondoftwo}%
\providecommand \bibfield  [0]{\@secondoftwo}%
\providecommand \translation [1]{[#1]}%
\providecommand \BibitemOpen [0]{}%
\providecommand \bibitemStop [0]{}%
\providecommand \bibitemNoStop [0]{.\EOS\space}%
\providecommand \EOS [0]{\spacefactor3000\relax}%
\providecommand \BibitemShut  [1]{\csname bibitem#1\endcsname}%
\let\auto@bib@innerbib\@empty
\bibitem [{\citenamefont {Barnett}(1915)}]{Barnett}%
  \BibitemOpen
  \bibfield  {author} {\bibinfo {author} {\bibfnamefont {S.~J.}\ \bibnamefont
  {Barnett}},\ }\href {https://link.aps.org/doi/10.1103/PhysRev.6.239}
  {\bibfield  {journal} {\bibinfo  {journal} {Phys. Rev.}\ }\textbf {\bibinfo
  {volume} {6}},\ \bibinfo {pages} {239} (\bibinfo {year} {1915})}\BibitemShut
  {NoStop}%
\bibitem [{\citenamefont {Barnett}(1935)}]{Barnett2}%
  \BibitemOpen
  \bibfield  {author} {\bibinfo {author} {\bibfnamefont {S.~J.}\ \bibnamefont
  {Barnett}},\ }\href {https://link.aps.org/doi/10.1103/RevModPhys.7.129}
  {\bibfield  {journal} {\bibinfo  {journal} {Rev. Mod. Phys.}\ }\textbf
  {\bibinfo {volume} {7}},\ \bibinfo {pages} {129} (\bibinfo {year}
  {1935})}\BibitemShut {NoStop}%
\bibitem [{\citenamefont {Einstein}\ and\ \citenamefont {de~Haas}(1915)}]{EdH}%
  \BibitemOpen
  \bibfield  {author} {\bibinfo {author} {\bibfnamefont {A.}~\bibnamefont
  {Einstein}}\ and\ \bibinfo {author} {\bibfnamefont {W.~J.}\ \bibnamefont
  {de~Haas}},\ }\href
  {https://www.scopus.com/record/display.uri?eid=2-s2.0-0013410258&origin=inward}
  {\bibfield  {journal} {\bibinfo  {journal} {Verh. Dtsch. Phys. Ges.}\
  }\textbf {\bibinfo {volume} {17}},\ \bibinfo {pages} {152} (\bibinfo {year}
  {1915})}\BibitemShut {NoStop}%
\bibitem [{\citenamefont {Ono}\ \emph {et~al.}(2015)\citenamefont {Ono},
  \citenamefont {Chudo}, \citenamefont {Harii}, \citenamefont {Okayasu},
  \citenamefont {Matsuo}, \citenamefont {Ieda}, \citenamefont {Takahashi},
  \citenamefont {Maekawa},\ and\ \citenamefont {Saitoh}}]{OnoBarnett}%
  \BibitemOpen
  \bibfield  {author} {\bibinfo {author} {\bibfnamefont {M.}~\bibnamefont
  {Ono}}, \bibinfo {author} {\bibfnamefont {H.}~\bibnamefont {Chudo}}, \bibinfo
  {author} {\bibfnamefont {K.}~\bibnamefont {Harii}}, \bibinfo {author}
  {\bibfnamefont {S.}~\bibnamefont {Okayasu}}, \bibinfo {author} {\bibfnamefont
  {M.}~\bibnamefont {Matsuo}}, \bibinfo {author} {\bibfnamefont
  {J.}~\bibnamefont {Ieda}}, \bibinfo {author} {\bibfnamefont {R.}~\bibnamefont
  {Takahashi}}, \bibinfo {author} {\bibfnamefont {S.}~\bibnamefont {Maekawa}},
  \ and\ \bibinfo {author} {\bibfnamefont {E.}~\bibnamefont {Saitoh}},\ }\href
  {https://link.aps.org/doi/10.1103/PhysRevB.92.174424} {\bibfield  {journal}
  {\bibinfo  {journal} {Phys. Rev. B}\ }\textbf {\bibinfo {volume} {92}},\
  \bibinfo {pages} {174424} (\bibinfo {year} {2015})}\BibitemShut {NoStop}%
\bibitem [{\citenamefont {Chudo}\ \emph {et~al.}(2014)\citenamefont {Chudo},
  \citenamefont {Ono}, \citenamefont {Harii}, \citenamefont {Matsuo},
  \citenamefont {Ieda}, \citenamefont {Haruki}, \citenamefont {Okayasu},
  \citenamefont {Maekawa}, \citenamefont {Yasuoka},\ and\ \citenamefont
  {Saitoh}}]{ChudoBarnett}%
  \BibitemOpen
  \bibfield  {author} {\bibinfo {author} {\bibfnamefont {H.}~\bibnamefont
  {Chudo}}, \bibinfo {author} {\bibfnamefont {M.}~\bibnamefont {Ono}}, \bibinfo
  {author} {\bibfnamefont {K.}~\bibnamefont {Harii}}, \bibinfo {author}
  {\bibfnamefont {M.}~\bibnamefont {Matsuo}}, \bibinfo {author} {\bibfnamefont
  {J.}~\bibnamefont {Ieda}}, \bibinfo {author} {\bibfnamefont {R.}~\bibnamefont
  {Haruki}}, \bibinfo {author} {\bibfnamefont {S.}~\bibnamefont {Okayasu}},
  \bibinfo {author} {\bibfnamefont {S.}~\bibnamefont {Maekawa}}, \bibinfo
  {author} {\bibfnamefont {H.}~\bibnamefont {Yasuoka}}, \ and\ \bibinfo
  {author} {\bibfnamefont {E.}~\bibnamefont {Saitoh}},\ }\href
  {https://iopscience.iop.org/article/10.7567/APEX.7.063004/meta} {\bibfield
  {journal} {\bibinfo  {journal} {Appl. Phys. Express}\ }\textbf {\bibinfo
  {volume} {7}},\ \bibinfo {pages} {063004} (\bibinfo {year}
  {2014})}\BibitemShut {NoStop}%
\bibitem [{\citenamefont {Arabgol}\ and\ \citenamefont
  {Sleator}(2019)}]{NuclearBarnett}%
  \BibitemOpen
  \bibfield  {author} {\bibinfo {author} {\bibfnamefont {M.}~\bibnamefont
  {Arabgol}}\ and\ \bibinfo {author} {\bibfnamefont {T.}~\bibnamefont
  {Sleator}},\ }\href {https://link.aps.org/doi/10.1103/PhysRevLett.122.177202}
  {\bibfield  {journal} {\bibinfo  {journal} {Phys. Rev. Lett.}\ }\textbf
  {\bibinfo {volume} {122}},\ \bibinfo {pages} {177202} (\bibinfo {year}
  {2019})}\BibitemShut {NoStop}%
\bibitem [{\citenamefont {Mukai}\ \emph {et~al.}(2014)\citenamefont {Mukai},
  \citenamefont {Hirori}, \citenamefont {Yamamoto}, \citenamefont {Kageyama},\
  and\ \citenamefont {Tanaka}}]{Mukai2014APL}%
  \BibitemOpen
  \bibfield  {author} {\bibinfo {author} {\bibfnamefont {Y.}~\bibnamefont
  {Mukai}}, \bibinfo {author} {\bibfnamefont {H.}~\bibnamefont {Hirori}},
  \bibinfo {author} {\bibfnamefont {T.}~\bibnamefont {Yamamoto}}, \bibinfo
  {author} {\bibfnamefont {H.}~\bibnamefont {Kageyama}}, \ and\ \bibinfo
  {author} {\bibfnamefont {K.}~\bibnamefont {Tanaka}},\ }\href {\doibase
  10.1063/1.4890475} {\bibfield  {journal} {\bibinfo  {journal} {Appl. Phys.
  Lett.}\ }\textbf {\bibinfo {volume} {105}},\ \bibinfo {pages} {022410}
  (\bibinfo {year} {2014})}\BibitemShut {NoStop}%
\bibitem [{\citenamefont {Bossini}\ \emph {et~al.}(2016)\citenamefont
  {Bossini}, \citenamefont {Belotelov}, \citenamefont {Zvezdin}, \citenamefont
  {Kalish},\ and\ \citenamefont {Kimel}}]{LaserPhotoExp}%
  \BibitemOpen
  \bibfield  {author} {\bibinfo {author} {\bibfnamefont {D.}~\bibnamefont
  {Bossini}}, \bibinfo {author} {\bibfnamefont {V.~I.}\ \bibnamefont
  {Belotelov}}, \bibinfo {author} {\bibfnamefont {A.~K.}\ \bibnamefont
  {Zvezdin}}, \bibinfo {author} {\bibfnamefont {A.~N.}\ \bibnamefont {Kalish}},
  \ and\ \bibinfo {author} {\bibfnamefont {A.~V.}\ \bibnamefont {Kimel}},\
  }\href {https://pubs.acs.org/doi/10.1021/acsphotonics.6b00107} {\bibfield
  {journal} {\bibinfo  {journal} {ACS Photon.}\ }\textbf {\bibinfo {volume}
  {3}},\ \bibinfo {pages} {1385} (\bibinfo {year} {2016})}\BibitemShut
  {NoStop}%
\bibitem [{\citenamefont {Ciappina}\ \emph {et~al.}(2017)\citenamefont
  {Ciappina}, \citenamefont {P.-Hernandez}, \citenamefont {Landsman},
  \citenamefont {Okell}, \citenamefont {Zherebtsov}, \citenamefont {Forg},
  \citenamefont {Schotz}, \citenamefont {Seiffert}, \citenamefont {Fennel},
  \citenamefont {Shaaran}, \citenamefont {Zimmermann}, \citenamefont {Chacon},
  \citenamefont {Guichard}, \citenamefont {Zair}, \citenamefont {Tisch},
  \citenamefont {Marangos}, \citenamefont {Witting}, \citenamefont {Braun},
  \citenamefont {Maier}, \citenamefont {Roso}, \citenamefont {Kruger},
  \citenamefont {Hommelhoff}, \citenamefont {Kling}, \citenamefont {Krausz},\
  and\ \citenamefont {Lewenstein}}]{Ciappina2017RepProgPhys}%
  \BibitemOpen
  \bibfield  {author} {\bibinfo {author} {\bibfnamefont {M.~F.}\ \bibnamefont
  {Ciappina}}, \bibinfo {author} {\bibfnamefont {J.~A.}\ \bibnamefont
  {P.-Hernandez}}, \bibinfo {author} {\bibfnamefont {A.~S.}\ \bibnamefont
  {Landsman}}, \bibinfo {author} {\bibfnamefont {W.~A.}\ \bibnamefont {Okell}},
  \bibinfo {author} {\bibfnamefont {S.}~\bibnamefont {Zherebtsov}}, \bibinfo
  {author} {\bibfnamefont {B.}~\bibnamefont {Forg}}, \bibinfo {author}
  {\bibfnamefont {J.}~\bibnamefont {Schotz}}, \bibinfo {author} {\bibfnamefont
  {L.}~\bibnamefont {Seiffert}}, \bibinfo {author} {\bibfnamefont
  {T.}~\bibnamefont {Fennel}}, \bibinfo {author} {\bibfnamefont
  {T.}~\bibnamefont {Shaaran}}, \bibinfo {author} {\bibfnamefont
  {T.}~\bibnamefont {Zimmermann}}, \bibinfo {author} {\bibfnamefont
  {A.}~\bibnamefont {Chacon}}, \bibinfo {author} {\bibfnamefont
  {R.}~\bibnamefont {Guichard}}, \bibinfo {author} {\bibfnamefont
  {A.}~\bibnamefont {Zair}}, \bibinfo {author} {\bibfnamefont {J.~W.~G.}\
  \bibnamefont {Tisch}}, \bibinfo {author} {\bibfnamefont {J.~P.}\ \bibnamefont
  {Marangos}}, \bibinfo {author} {\bibfnamefont {T.}~\bibnamefont {Witting}},
  \bibinfo {author} {\bibfnamefont {A.}~\bibnamefont {Braun}}, \bibinfo
  {author} {\bibfnamefont {S.~A.}\ \bibnamefont {Maier}}, \bibinfo {author}
  {\bibfnamefont {L.}~\bibnamefont {Roso}}, \bibinfo {author} {\bibfnamefont
  {M.}~\bibnamefont {Kruger}}, \bibinfo {author} {\bibfnamefont
  {P.}~\bibnamefont {Hommelhoff}}, \bibinfo {author} {\bibfnamefont {M.~F.}\
  \bibnamefont {Kling}}, \bibinfo {author} {\bibfnamefont {F.}~\bibnamefont
  {Krausz}}, \ and\ \bibinfo {author} {\bibfnamefont {M.}~\bibnamefont
  {Lewenstein}},\ }\href
  {https://iopscience.iop.org/article/10.1088/1361-6633/aa574e} {\bibfield
  {journal} {\bibinfo  {journal} {Rep. Prog. Phys.}\ }\textbf {\bibinfo
  {volume} {80}},\ \bibinfo {pages} {054401} (\bibinfo {year}
  {2017})}\BibitemShut {NoStop}%
\bibitem [{\citenamefont {Arikawa}\ \emph {et~al.}(2017)\citenamefont
  {Arikawa}, \citenamefont {Morimoto},\ and\ \citenamefont
  {Tanaka}}]{LaserPhotoExp3}%
  \BibitemOpen
  \bibfield  {author} {\bibinfo {author} {\bibfnamefont {T.}~\bibnamefont
  {Arikawa}}, \bibinfo {author} {\bibfnamefont {S.}~\bibnamefont {Morimoto}}, \
  and\ \bibinfo {author} {\bibfnamefont {K.}~\bibnamefont {Tanaka}},\ }\href
  {https://www.osapublishing.org/oe/abstract.cfm?uri=oe-25-12-13728} {\bibfield
   {journal} {\bibinfo  {journal} {Opt. Express}\ }\textbf {\bibinfo {volume}
  {25}},\ \bibinfo {pages} {13728} (\bibinfo {year} {2017})}\BibitemShut
  {NoStop}%
\bibitem [{\citenamefont {Stanciu}\ \emph {et~al.}(2007)\citenamefont
  {Stanciu}, \citenamefont {Hansteen}, \citenamefont {Kimel}, \citenamefont
  {Kirilyuk}, \citenamefont {Tsukamoto}, \citenamefont {Itoh},\ and\
  \citenamefont {Rasing}}]{OtherOpticalBarnett}%
  \BibitemOpen
  \bibfield  {author} {\bibinfo {author} {\bibfnamefont {C.~D.}\ \bibnamefont
  {Stanciu}}, \bibinfo {author} {\bibfnamefont {F.}~\bibnamefont {Hansteen}},
  \bibinfo {author} {\bibfnamefont {A.~V.}\ \bibnamefont {Kimel}}, \bibinfo
  {author} {\bibfnamefont {A.}~\bibnamefont {Kirilyuk}}, \bibinfo {author}
  {\bibfnamefont {A.}~\bibnamefont {Tsukamoto}}, \bibinfo {author}
  {\bibfnamefont {A.}~\bibnamefont {Itoh}}, \ and\ \bibinfo {author}
  {\bibfnamefont {T.}~\bibnamefont {Rasing}},\ }\href
  {https://link.aps.org/doi/10.1103/PhysRevLett.99.047601} {\bibfield
  {journal} {\bibinfo  {journal} {Phys. Rev. Lett.}\ }\textbf {\bibinfo
  {volume} {99}},\ \bibinfo {pages} {047601} (\bibinfo {year}
  {2007})}\BibitemShut {NoStop}%
\bibitem [{\citenamefont {Rebei}\ and\ \citenamefont
  {Hohlfeld}(2008{\natexlab{a}})}]{OtherOpticalBarnett2}%
  \BibitemOpen
  \bibfield  {author} {\bibinfo {author} {\bibfnamefont {A.}~\bibnamefont
  {Rebei}}\ and\ \bibinfo {author} {\bibfnamefont {J.}~\bibnamefont
  {Hohlfeld}},\ }\href
  {https://www.sciencedirect.com/science/article/pii/S0375960107015198}
  {\bibfield  {journal} {\bibinfo  {journal} {Phys. Lett. A}\ }\textbf
  {\bibinfo {volume} {372}},\ \bibinfo {pages} {1915} (\bibinfo {year}
  {2008}{\natexlab{a}})}\BibitemShut {NoStop}%
\bibitem [{\citenamefont {Rebei}\ and\ \citenamefont
  {Hohlfeld}(2008{\natexlab{b}})}]{OtherOpticalBarnett3}%
  \BibitemOpen
  \bibfield  {author} {\bibinfo {author} {\bibfnamefont {A.}~\bibnamefont
  {Rebei}}\ and\ \bibinfo {author} {\bibfnamefont {J.}~\bibnamefont
  {Hohlfeld}},\ }\href {https://aip.scitation.org/doi/full/10.1063/1.2837667}
  {\bibfield  {journal} {\bibinfo  {journal} {J. Appl. Phys.}\ }\textbf
  {\bibinfo {volume} {103}},\ \bibinfo {pages} {07B118} (\bibinfo {year}
  {2008}{\natexlab{b}})}\BibitemShut {NoStop}%
\bibitem [{\citenamefont {Kirilyuk}\ \emph {et~al.}(2010)\citenamefont
  {Kirilyuk}, \citenamefont {Kimel},\ and\ \citenamefont {Rasing}}]{Kirilyuk}%
  \BibitemOpen
  \bibfield  {author} {\bibinfo {author} {\bibfnamefont {A.}~\bibnamefont
  {Kirilyuk}}, \bibinfo {author} {\bibfnamefont {A.~V.}\ \bibnamefont {Kimel}},
  \ and\ \bibinfo {author} {\bibfnamefont {T.}~\bibnamefont {Rasing}},\ }\href
  {\doibase 10.1103/RevModPhys.82.2731} {\bibfield  {journal} {\bibinfo
  {journal} {Rev. Mod. Phys.}\ }\textbf {\bibinfo {volume} {82}},\ \bibinfo
  {pages} {2731} (\bibinfo {year} {2010})}\BibitemShut {NoStop}%
\bibitem [{\citenamefont {Kimel}\ \emph {et~al.}(2005)\citenamefont {Kimel},
  \citenamefont {Kirilyuk}, \citenamefont {Usachev}, \citenamefont {Pisarev},
  \citenamefont {Balbashov},\ and\ \citenamefont
  {Rasing}}]{KimelNatureIFaraday}%
  \BibitemOpen
  \bibfield  {author} {\bibinfo {author} {\bibfnamefont {A.~V.}\ \bibnamefont
  {Kimel}}, \bibinfo {author} {\bibfnamefont {A.}~\bibnamefont {Kirilyuk}},
  \bibinfo {author} {\bibfnamefont {P.~A.}\ \bibnamefont {Usachev}}, \bibinfo
  {author} {\bibfnamefont {R.~V.}\ \bibnamefont {Pisarev}}, \bibinfo {author}
  {\bibfnamefont {A.~M.}\ \bibnamefont {Balbashov}}, \ and\ \bibinfo {author}
  {\bibfnamefont {T.}~\bibnamefont {Rasing}},\ }\href
  {https://www.nature.com/articles/nature03564} {\bibfield  {journal} {\bibinfo
   {journal} {Nature}\ }\textbf {\bibinfo {volume} {435}},\ \bibinfo {pages}
  {655} (\bibinfo {year} {2005})}\BibitemShut {NoStop}%
\bibitem [{\citenamefont {Osada}\ \emph {et~al.}(2016)\citenamefont {Osada},
  \citenamefont {Hisatomi}, \citenamefont {Noguchi}, \citenamefont {Tabuchi},
  \citenamefont {Yamazaki}, \citenamefont {Usami}, \citenamefont {Sadgrove},
  \citenamefont {Yalla}, \citenamefont {Nomura},\ and\ \citenamefont
  {Nakamura}}]{OptomagnonicsCavity}%
  \BibitemOpen
  \bibfield  {author} {\bibinfo {author} {\bibfnamefont {A.}~\bibnamefont
  {Osada}}, \bibinfo {author} {\bibfnamefont {R.}~\bibnamefont {Hisatomi}},
  \bibinfo {author} {\bibfnamefont {A.}~\bibnamefont {Noguchi}}, \bibinfo
  {author} {\bibfnamefont {Y.}~\bibnamefont {Tabuchi}}, \bibinfo {author}
  {\bibfnamefont {R.}~\bibnamefont {Yamazaki}}, \bibinfo {author}
  {\bibfnamefont {K.}~\bibnamefont {Usami}}, \bibinfo {author} {\bibfnamefont
  {M.}~\bibnamefont {Sadgrove}}, \bibinfo {author} {\bibfnamefont
  {R.}~\bibnamefont {Yalla}}, \bibinfo {author} {\bibfnamefont
  {M.}~\bibnamefont {Nomura}}, \ and\ \bibinfo {author} {\bibfnamefont
  {Y.}~\bibnamefont {Nakamura}},\ }\href
  {https://link.aps.org/doi/10.1103/PhysRevLett.116.223601} {\bibfield
  {journal} {\bibinfo  {journal} {Phys. Rev. Lett.}\ }\textbf {\bibinfo
  {volume} {116}},\ \bibinfo {pages} {223601} (\bibinfo {year}
  {2016})}\BibitemShut {NoStop}%
\bibitem [{\citenamefont {Liu}\ \emph {et~al.}(2016)\citenamefont {Liu},
  \citenamefont {Zhang}, \citenamefont {Tang},\ and\ \citenamefont
  {Flatt$\acute{\rm{e}}$}}]{OptomagnonicsCavity2}%
  \BibitemOpen
  \bibfield  {author} {\bibinfo {author} {\bibfnamefont {T.}~\bibnamefont
  {Liu}}, \bibinfo {author} {\bibfnamefont {X.}~\bibnamefont {Zhang}}, \bibinfo
  {author} {\bibfnamefont {H.~X.}\ \bibnamefont {Tang}}, \ and\ \bibinfo
  {author} {\bibfnamefont {M.~E.}\ \bibnamefont {Flatt$\acute{\rm{e}}$}},\
  }\href {https://link.aps.org/doi/10.1103/PhysRevB.94.060405} {\bibfield
  {journal} {\bibinfo  {journal} {Phys. Rev. B}\ }\textbf {\bibinfo {volume}
  {94}},\ \bibinfo {pages} {060405(R)} (\bibinfo {year} {2016})}\BibitemShut
  {NoStop}%
\bibitem [{\citenamefont {Kusminskiy}\ \emph {et~al.}(2016)\citenamefont
  {Kusminskiy}, \citenamefont {Tang},\ and\ \citenamefont
  {Marquardt}}]{OptomagnonicsCavity3}%
  \BibitemOpen
  \bibfield  {author} {\bibinfo {author} {\bibfnamefont {S.~V.}\ \bibnamefont
  {Kusminskiy}}, \bibinfo {author} {\bibfnamefont {H.~X.}\ \bibnamefont
  {Tang}}, \ and\ \bibinfo {author} {\bibfnamefont {F.}~\bibnamefont
  {Marquardt}},\ }\href {https://link.aps.org/doi/10.1103/PhysRevA.94.033821}
  {\bibfield  {journal} {\bibinfo  {journal} {Phys. Rev. A}\ }\textbf {\bibinfo
  {volume} {94}},\ \bibinfo {pages} {033821} (\bibinfo {year}
  {2016})}\BibitemShut {NoStop}%
\bibitem [{\citenamefont {Nakata}\ \emph
  {et~al.}(2017{\natexlab{a}})\citenamefont {Nakata}, \citenamefont {Simon},\
  and\ \citenamefont {Loss}}]{ReviewMagnon}%
  \BibitemOpen
  \bibfield  {author} {\bibinfo {author} {\bibfnamefont {K.}~\bibnamefont
  {Nakata}}, \bibinfo {author} {\bibfnamefont {P.}~\bibnamefont {Simon}}, \
  and\ \bibinfo {author} {\bibfnamefont {D.}~\bibnamefont {Loss}},\ }\href
  {https://iopscience.iop.org/article/10.1088/1361-6463/aa5b09} {\bibfield
  {journal} {\bibinfo  {journal} {J. Phys. D: Appl. Phys.}\ }\textbf {\bibinfo
  {volume} {50}},\ \bibinfo {pages} {114004} (\bibinfo {year}
  {2017}{\natexlab{a}})}\BibitemShut {NoStop}%
\bibitem [{\citenamefont {Chumak}\ \emph {et~al.}(2015)\citenamefont {Chumak},
  \citenamefont {Vasyuchka}, \citenamefont {Serga},\ and\ \citenamefont
  {Hillebrands}}]{MagnonSpintronics}%
  \BibitemOpen
  \bibfield  {author} {\bibinfo {author} {\bibfnamefont {A.~V.}\ \bibnamefont
  {Chumak}}, \bibinfo {author} {\bibfnamefont {V.~I.}\ \bibnamefont
  {Vasyuchka}}, \bibinfo {author} {\bibfnamefont {A.~A.}\ \bibnamefont
  {Serga}}, \ and\ \bibinfo {author} {\bibfnamefont {B.}~\bibnamefont
  {Hillebrands}},\ }\href {https://www.nature.com/articles/nphys3347}
  {\bibfield  {journal} {\bibinfo  {journal} {Nat. Phys.}\ }\textbf {\bibinfo
  {volume} {11}},\ \bibinfo {pages} {453} (\bibinfo {year} {2015})}\BibitemShut
  {NoStop}%
\bibitem [{\citenamefont {Pershan}\ \emph {et~al.}(1966)\citenamefont
  {Pershan}, \citenamefont {van~der Ziel},\ and\ \citenamefont
  {Malmstrom}}]{IFE}%
  \BibitemOpen
  \bibfield  {author} {\bibinfo {author} {\bibfnamefont {P.~S.}\ \bibnamefont
  {Pershan}}, \bibinfo {author} {\bibfnamefont {J.~P.}\ \bibnamefont {van~der
  Ziel}}, \ and\ \bibinfo {author} {\bibfnamefont {L.~D.}\ \bibnamefont
  {Malmstrom}},\ }\href {https://link.aps.org/doi/10.1103/PhysRev.143.574}
  {\bibfield  {journal} {\bibinfo  {journal} {Phys. Rev.}\ }\textbf {\bibinfo
  {volume} {143}},\ \bibinfo {pages} {574} (\bibinfo {year}
  {1966})}\BibitemShut {NoStop}%
\bibitem [{\citenamefont {Takayoshi}\ \emph
  {et~al.}(2014{\natexlab{a}})\citenamefont {Takayoshi}, \citenamefont {Sato},\
  and\ \citenamefont {Oka}}]{FloquetST2}%
  \BibitemOpen
  \bibfield  {author} {\bibinfo {author} {\bibfnamefont {S.}~\bibnamefont
  {Takayoshi}}, \bibinfo {author} {\bibfnamefont {M.}~\bibnamefont {Sato}}, \
  and\ \bibinfo {author} {\bibfnamefont {T.}~\bibnamefont {Oka}},\ }\href
  {https://link.aps.org/doi/10.1103/PhysRevB.90.214413} {\bibfield  {journal}
  {\bibinfo  {journal} {Phys. Rev. B}\ }\textbf {\bibinfo {volume} {90}},\
  \bibinfo {pages} {214413} (\bibinfo {year} {2014}{\natexlab{a}})}\BibitemShut
  {NoStop}%
\bibitem [{\citenamefont {Takayoshi}\ \emph
  {et~al.}(2014{\natexlab{b}})\citenamefont {Takayoshi}, \citenamefont {Aoki},\
  and\ \citenamefont {Oka}}]{FloquetST}%
  \BibitemOpen
  \bibfield  {author} {\bibinfo {author} {\bibfnamefont {S.}~\bibnamefont
  {Takayoshi}}, \bibinfo {author} {\bibfnamefont {H.}~\bibnamefont {Aoki}}, \
  and\ \bibinfo {author} {\bibfnamefont {T.}~\bibnamefont {Oka}},\ }\href
  {https://link.aps.org/doi/10.1103/PhysRevB.90.085150} {\bibfield  {journal}
  {\bibinfo  {journal} {Phys. Rev. B}\ }\textbf {\bibinfo {volume} {90}},\
  \bibinfo {pages} {085150} (\bibinfo {year} {2014}{\natexlab{b}})}\BibitemShut
  {NoStop}%
\bibitem [{\citenamefont {de~Oliveira}\ and\ \citenamefont
  {Tiomno}(1962)}]{SRC}%
  \BibitemOpen
  \bibfield  {author} {\bibinfo {author} {\bibfnamefont {C.~G.}\ \bibnamefont
  {de~Oliveira}}\ and\ \bibinfo {author} {\bibfnamefont {J.}~\bibnamefont
  {Tiomno}},\ }\href {https://link.springer.com/article/10.1007/BF02816716}
  {\bibfield  {journal} {\bibinfo  {journal} {Nuovo Cimento}\ }\textbf
  {\bibinfo {volume} {24}},\ \bibinfo {pages} {672} (\bibinfo {year}
  {1962})}\BibitemShut {NoStop}%
\bibitem [{\citenamefont {Mashhoon}(1988)}]{SRC2}%
  \BibitemOpen
  \bibfield  {author} {\bibinfo {author} {\bibfnamefont {B.}~\bibnamefont
  {Mashhoon}},\ }\href {https://link.aps.org/doi/10.1103/PhysRevLett.61.2639}
  {\bibfield  {journal} {\bibinfo  {journal} {Phys. Rev. Lett.}\ }\textbf
  {\bibinfo {volume} {61}},\ \bibinfo {pages} {2639} (\bibinfo {year}
  {1988})}\BibitemShut {NoStop}%
\bibitem [{\citenamefont {Hehl}\ and\ \citenamefont {Ni}(1990)}]{SRC3}%
  \BibitemOpen
  \bibfield  {author} {\bibinfo {author} {\bibfnamefont {F.~W.}\ \bibnamefont
  {Hehl}}\ and\ \bibinfo {author} {\bibfnamefont {W.-T.}\ \bibnamefont {Ni}},\
  }\href {https://link.aps.org/doi/10.1103/PhysRevD.42.2045} {\bibfield
  {journal} {\bibinfo  {journal} {Phys. Rev. D}\ }\textbf {\bibinfo {volume}
  {42}},\ \bibinfo {pages} {2045} (\bibinfo {year} {1990})}\BibitemShut
  {NoStop}%
\bibitem [{\citenamefont {Matsuo}\ \emph
  {et~al.}(2011{\natexlab{a}})\citenamefont {Matsuo}, \citenamefont {Ieda},
  \citenamefont {Saitoh},\ and\ \citenamefont {Maekawa}}]{SRCmm}%
  \BibitemOpen
  \bibfield  {author} {\bibinfo {author} {\bibfnamefont {M.}~\bibnamefont
  {Matsuo}}, \bibinfo {author} {\bibfnamefont {J.}~\bibnamefont {Ieda}},
  \bibinfo {author} {\bibfnamefont {E.}~\bibnamefont {Saitoh}}, \ and\ \bibinfo
  {author} {\bibfnamefont {S.}~\bibnamefont {Maekawa}},\ }\href
  {https://link.aps.org/doi/10.1103/PhysRevLett.106.076601} {\bibfield
  {journal} {\bibinfo  {journal} {Phys. Rev. Lett.}\ }\textbf {\bibinfo
  {volume} {106}},\ \bibinfo {pages} {076601} (\bibinfo {year}
  {2011}{\natexlab{a}})}\BibitemShut {NoStop}%
\bibitem [{\citenamefont {Matsuo}\ \emph
  {et~al.}(2011{\natexlab{b}})\citenamefont {Matsuo}, \citenamefont {Ieda},
  \citenamefont {Saitoh},\ and\ \citenamefont {Maekawa}}]{SRCmm2}%
  \BibitemOpen
  \bibfield  {author} {\bibinfo {author} {\bibfnamefont {M.}~\bibnamefont
  {Matsuo}}, \bibinfo {author} {\bibfnamefont {J.}~\bibnamefont {Ieda}},
  \bibinfo {author} {\bibfnamefont {E.}~\bibnamefont {Saitoh}}, \ and\ \bibinfo
  {author} {\bibfnamefont {S.}~\bibnamefont {Maekawa}},\ }\href
  {https://link.aps.org/doi/10.1103/PhysRevB.84.104410} {\bibfield  {journal}
  {\bibinfo  {journal} {Phys. Rev. B}\ }\textbf {\bibinfo {volume} {84}},\
  \bibinfo {pages} {104410} (\bibinfo {year} {2011}{\natexlab{b}})}\BibitemShut
  {NoStop}%
\bibitem [{\citenamefont {Matsuo}\ \emph
  {et~al.}(2013{\natexlab{a}})\citenamefont {Matsuo}, \citenamefont {Ieda},\
  and\ \citenamefont {Maekawa}}]{SRCmm3}%
  \BibitemOpen
  \bibfield  {author} {\bibinfo {author} {\bibfnamefont {M.}~\bibnamefont
  {Matsuo}}, \bibinfo {author} {\bibfnamefont {J.}~\bibnamefont {Ieda}}, \ and\
  \bibinfo {author} {\bibfnamefont {S.}~\bibnamefont {Maekawa}},\ }\href
  {https://link.aps.org/doi/10.1103/PhysRevB.87.115301} {\bibfield  {journal}
  {\bibinfo  {journal} {Phys. Rev. B}\ }\textbf {\bibinfo {volume} {87}},\
  \bibinfo {pages} {115301} (\bibinfo {year} {2013}{\natexlab{a}})}\BibitemShut
  {NoStop}%
\bibitem [{\citenamefont {Matsuo}\ \emph
  {et~al.}(2013{\natexlab{b}})\citenamefont {Matsuo}, \citenamefont {Ieda},
  \citenamefont {Harii}, \citenamefont {Saitoh},\ and\ \citenamefont
  {Maekawa}}]{SRCmm4}%
  \BibitemOpen
  \bibfield  {author} {\bibinfo {author} {\bibfnamefont {M.}~\bibnamefont
  {Matsuo}}, \bibinfo {author} {\bibfnamefont {J.}~\bibnamefont {Ieda}},
  \bibinfo {author} {\bibfnamefont {K.}~\bibnamefont {Harii}}, \bibinfo
  {author} {\bibfnamefont {E.}~\bibnamefont {Saitoh}}, \ and\ \bibinfo {author}
  {\bibfnamefont {S.}~\bibnamefont {Maekawa}},\ }\href
  {https://link.aps.org/doi/10.1103/PhysRevB.87.180402} {\bibfield  {journal}
  {\bibinfo  {journal} {Phys. Rev. B}\ }\textbf {\bibinfo {volume} {87}},\
  \bibinfo {pages} {180402(R)} (\bibinfo {year}
  {2013}{\natexlab{b}})}\BibitemShut {NoStop}%
\bibitem [{\citenamefont {Matsuo}\ \emph {et~al.}(2017)\citenamefont {Matsuo},
  \citenamefont {Saitoh},\ and\ \citenamefont
  {Maekawa}}]{ReviewSpinMechatronics}%
  \BibitemOpen
  \bibfield  {author} {\bibinfo {author} {\bibfnamefont {M.}~\bibnamefont
  {Matsuo}}, \bibinfo {author} {\bibfnamefont {E.}~\bibnamefont {Saitoh}}, \
  and\ \bibinfo {author} {\bibfnamefont {S.}~\bibnamefont {Maekawa}},\ }\href
  {https://journals.jps.jp/doi/full/10.7566/JPSJ.86.011011} {\bibfield
  {journal} {\bibinfo  {journal} {J. Phys. Soc. Jpn.}\ }\textbf {\bibinfo
  {volume} {86}},\ \bibinfo {pages} {011011} (\bibinfo {year}
  {2017})}\BibitemShut {NoStop}%
\bibitem [{\citenamefont {Ohnuma}\ \emph {et~al.}(2013)\citenamefont {Ohnuma},
  \citenamefont {Adachi}, \citenamefont {Saitoh},\ and\ \citenamefont
  {Maekawa}}]{ohnuma}%
  \BibitemOpen
  \bibfield  {author} {\bibinfo {author} {\bibfnamefont {Y.}~\bibnamefont
  {Ohnuma}}, \bibinfo {author} {\bibfnamefont {H.}~\bibnamefont {Adachi}},
  \bibinfo {author} {\bibfnamefont {E.}~\bibnamefont {Saitoh}}, \ and\ \bibinfo
  {author} {\bibfnamefont {S.}~\bibnamefont {Maekawa}},\ }\href
  {https://link.aps.org/doi/10.1103/PhysRevB.87.014423} {\bibfield  {journal}
  {\bibinfo  {journal} {Phys. Rev. B}\ }\textbf {\bibinfo {volume} {87}},\
  \bibinfo {pages} {014423} (\bibinfo {year} {2013})}\BibitemShut {NoStop}%
\bibitem [{\citenamefont {Chikazumi}(1997)}]{FerriOhnuma}%
  \BibitemOpen
  \bibfield  {author} {\bibinfo {author} {\bibfnamefont {S.}~\bibnamefont
  {Chikazumi}},\ }\href@noop {} {\emph {\bibinfo {title} {Physics of
  Ferromagnetism}}}\ (\bibinfo  {publisher} {Oxford Science, New York},\
  \bibinfo {year} {1997})\BibitemShut {NoStop}%
\bibitem [{\citenamefont {Pearson}(1962)}]{FerriOhnuma2}%
  \BibitemOpen
  \bibfield  {author} {\bibinfo {author} {\bibfnamefont {R.~F.}\ \bibnamefont
  {Pearson}},\ }\href {https://aip.scitation.org/doi/10.1063/1.1728675}
  {\bibfield  {journal} {\bibinfo  {journal} {J. Appl. Phys.}\ }\textbf
  {\bibinfo {volume} {33}},\ \bibinfo {pages} {1236} (\bibinfo {year}
  {1962})}\BibitemShut {NoStop}%
\bibitem [{\citenamefont {Sato}\ \emph {et~al.}(2013)\citenamefont {Sato},
  \citenamefont {Higuchi}, \citenamefont {Kanda}, \citenamefont {Konishi},
  \citenamefont {Yoshioka}, \citenamefont {Suzuki}, \citenamefont {Misawa},\
  and\ \citenamefont {K.-Gonokami}}]{chirping}%
  \BibitemOpen
  \bibfield  {author} {\bibinfo {author} {\bibfnamefont {M.}~\bibnamefont
  {Sato}}, \bibinfo {author} {\bibfnamefont {T.}~\bibnamefont {Higuchi}},
  \bibinfo {author} {\bibfnamefont {N.}~\bibnamefont {Kanda}}, \bibinfo
  {author} {\bibfnamefont {K.}~\bibnamefont {Konishi}}, \bibinfo {author}
  {\bibfnamefont {K.}~\bibnamefont {Yoshioka}}, \bibinfo {author}
  {\bibfnamefont {T.}~\bibnamefont {Suzuki}}, \bibinfo {author} {\bibfnamefont
  {K.}~\bibnamefont {Misawa}}, \ and\ \bibinfo {author} {\bibfnamefont
  {M.}~\bibnamefont {K.-Gonokami}},\ }\href
  {https://www.nature.com/articles/nphoton.2013.213} {\bibfield  {journal}
  {\bibinfo  {journal} {Nat. Photon.}\ }\textbf {\bibinfo {volume} {7}},\
  \bibinfo {pages} {724} (\bibinfo {year} {2013})}\BibitemShut {NoStop}%
\bibitem [{\citenamefont {Kamada}\ \emph {et~al.}(2013)\citenamefont {Kamada},
  \citenamefont {Murata},\ and\ \citenamefont {Aoki}}]{chirping2}%
  \BibitemOpen
  \bibfield  {author} {\bibinfo {author} {\bibfnamefont {S.}~\bibnamefont
  {Kamada}}, \bibinfo {author} {\bibfnamefont {S.}~\bibnamefont {Murata}}, \
  and\ \bibinfo {author} {\bibfnamefont {T.}~\bibnamefont {Aoki}},\ }\href
  {https://iopscience.iop.org/article/10.7567/APEX.6.032701} {\bibfield
  {journal} {\bibinfo  {journal} {Appl. Phys. Express}\ }\textbf {\bibinfo
  {volume} {6}},\ \bibinfo {pages} {032701} (\bibinfo {year}
  {2013})}\BibitemShut {NoStop}%
\bibitem [{\citenamefont {Chow}\ and\ \citenamefont
  {Keffer}(1974)}]{SpinFlopAF}%
  \BibitemOpen
  \bibfield  {author} {\bibinfo {author} {\bibfnamefont {H.}~\bibnamefont
  {Chow}}\ and\ \bibinfo {author} {\bibfnamefont {F.}~\bibnamefont {Keffer}},\
  }\href {\doibase 10.1103/PhysRevB.10.243} {\bibfield  {journal} {\bibinfo
  {journal} {Phys. Rev. B}\ }\textbf {\bibinfo {volume} {10}},\ \bibinfo
  {pages} {243} (\bibinfo {year} {1974})}\BibitemShut {NoStop}%
\bibitem [{\citenamefont {Clark}\ and\ \citenamefont
  {Callen}(1968)}]{SpinFlopTra}%
  \BibitemOpen
  \bibfield  {author} {\bibinfo {author} {\bibfnamefont {A.~E.}\ \bibnamefont
  {Clark}}\ and\ \bibinfo {author} {\bibfnamefont {E.}~\bibnamefont {Callen}},\
  }\href {https://aip.scitation.org/doi/10.1063/1.1656100} {\bibfield
  {journal} {\bibinfo  {journal} {J. Appl. Phys.}\ }\textbf {\bibinfo {volume}
  {39}},\ \bibinfo {pages} {5972} (\bibinfo {year} {1968})}\BibitemShut
  {NoStop}%
\bibitem [{\citenamefont {Nakata}\ \emph
  {et~al.}(2017{\natexlab{b}})\citenamefont {Nakata}, \citenamefont {Kim},
  \citenamefont {Klinovaja},\ and\ \citenamefont {Loss}}]{KSJD}%
  \BibitemOpen
  \bibfield  {author} {\bibinfo {author} {\bibfnamefont {K.}~\bibnamefont
  {Nakata}}, \bibinfo {author} {\bibfnamefont {S.~K.}\ \bibnamefont {Kim}},
  \bibinfo {author} {\bibfnamefont {J.}~\bibnamefont {Klinovaja}}, \ and\
  \bibinfo {author} {\bibfnamefont {D.}~\bibnamefont {Loss}},\ }\href
  {https://link.aps.org/doi/10.1103/PhysRevB.96.224414} {\bibfield  {journal}
  {\bibinfo  {journal} {Phys. Rev. B}\ }\textbf {\bibinfo {volume} {96}},\
  \bibinfo {pages} {224414} (\bibinfo {year} {2017}{\natexlab{b}})}\BibitemShut
  {NoStop}%
\bibitem [{Note1()}]{Note1}%
  \BibitemOpen
  \bibinfo {note} {The total number of magnons in the system is bounded by a
  hard-core interaction between magnons~\cite
  {oshikawa,TotsukaBEC2,Giamarchi2008NatPhys} arising from the higher order
  term in the spin wave theory. We neglect it for simplicity in Eqs.~\protect
  \textup {\hbox {\mathsurround \z@ \protect \normalfont (\ignorespaces \ref
  {eq:Hamil_MagBEC1}\unskip \@@italiccorr )}} and \protect \textup {\hbox
  {\mathsurround \z@ \protect \normalfont (\ignorespaces \ref
  {eq:Hamil_MagBEC2}\unskip \@@italiccorr )}}. Thereby the magnon BEC is stable
  in the system with a finite spin length.}\BibitemShut {Stop}%
\bibitem [{\citenamefont {Takayoshi}\ \emph {et~al.}(2019)\citenamefont
  {Takayoshi}, \citenamefont {Murakami},\ and\ \citenamefont
  {Werner}}]{Magdynam}%
  \BibitemOpen
  \bibfield  {author} {\bibinfo {author} {\bibfnamefont {S.}~\bibnamefont
  {Takayoshi}}, \bibinfo {author} {\bibfnamefont {Y.}~\bibnamefont {Murakami}},
  \ and\ \bibinfo {author} {\bibfnamefont {P.}~\bibnamefont {Werner}},\ }\href
  {\doibase 10.1103/PhysRevB.99.184303} {\bibfield  {journal} {\bibinfo
  {journal} {Phys. Rev. B}\ }\textbf {\bibinfo {volume} {99}},\ \bibinfo
  {pages} {184303} (\bibinfo {year} {2019})}\BibitemShut {NoStop}%
\bibitem [{Note2()}]{Note2}%
  \BibitemOpen
  \bibinfo {note} {Note that the quasiequilibrium magnon BEC reported in
  Ref.~\cite {demokritov} is experimentally realized by magnon injection
  through microwave pumping in the GHz regime.}\BibitemShut {Stop}%
\bibitem [{\citenamefont {Lee}\ \emph {et~al.}(2006)\citenamefont {Lee},
  \citenamefont {Hahn},\ and\ \citenamefont {Clarke}}]{LeeHahn}%
  \BibitemOpen
  \bibfield  {author} {\bibinfo {author} {\bibfnamefont {S.-K.}\ \bibnamefont
  {Lee}}, \bibinfo {author} {\bibfnamefont {E.~L.}\ \bibnamefont {Hahn}}, \
  and\ \bibinfo {author} {\bibfnamefont {J.}~\bibnamefont {Clarke}},\ }\href
  {\doibase 10.1103/PhysRevLett.96.257601} {\bibfield  {journal} {\bibinfo
  {journal} {Phys. Rev. Lett.}\ }\textbf {\bibinfo {volume} {96}},\ \bibinfo
  {pages} {257601} (\bibinfo {year} {2006})}\BibitemShut {NoStop}%
\bibitem [{\citenamefont {Qiu}\ and\ \citenamefont {Bader}(2000)}]{KerrEffect}%
  \BibitemOpen
  \bibfield  {author} {\bibinfo {author} {\bibfnamefont {Z.~Q.}\ \bibnamefont
  {Qiu}}\ and\ \bibinfo {author} {\bibfnamefont {S.~D.}\ \bibnamefont
  {Bader}},\ }\href {https://aip.scitation.org/doi/10.1063/1.1150496}
  {\bibfield  {journal} {\bibinfo  {journal} {Rev. Sci. Instrum.}\ }\textbf
  {\bibinfo {volume} {71}},\ \bibinfo {pages} {1243} (\bibinfo {year}
  {2000})}\BibitemShut {NoStop}%
\bibitem [{\citenamefont {Demokritov}\ \emph {et~al.}(2006)\citenamefont
  {Demokritov}, \citenamefont {Demidov}, \citenamefont {Dzyapko}, \citenamefont
  {Melkov}, \citenamefont {Serga}, \citenamefont {Hillebrands},\ and\
  \citenamefont {Slavin}}]{demokritov}%
  \BibitemOpen
  \bibfield  {author} {\bibinfo {author} {\bibfnamefont {S.~O.}\ \bibnamefont
  {Demokritov}}, \bibinfo {author} {\bibfnamefont {V.~E.}\ \bibnamefont
  {Demidov}}, \bibinfo {author} {\bibfnamefont {O.}~\bibnamefont {Dzyapko}},
  \bibinfo {author} {\bibfnamefont {G.~A.}\ \bibnamefont {Melkov}}, \bibinfo
  {author} {\bibfnamefont {A.~A.}\ \bibnamefont {Serga}}, \bibinfo {author}
  {\bibfnamefont {B.}~\bibnamefont {Hillebrands}}, \ and\ \bibinfo {author}
  {\bibfnamefont {A.~N.}\ \bibnamefont {Slavin}},\ }\href
  {https://www.nature.com/articles/nature05117} {\bibfield  {journal} {\bibinfo
   {journal} {Nature (London)}\ }\textbf {\bibinfo {volume} {443}},\ \bibinfo
  {pages} {430} (\bibinfo {year} {2006})}\BibitemShut {NoStop}%
\bibitem [{\citenamefont {Yamaguchi}\ \emph {et~al.}(2013)\citenamefont
  {Yamaguchi}, \citenamefont {Kurihara}, \citenamefont {Minami}, \citenamefont
  {Nakajima},\ and\ \citenamefont {Suemoto}}]{THzspectroscopy}%
  \BibitemOpen
  \bibfield  {author} {\bibinfo {author} {\bibfnamefont {K.}~\bibnamefont
  {Yamaguchi}}, \bibinfo {author} {\bibfnamefont {T.}~\bibnamefont {Kurihara}},
  \bibinfo {author} {\bibfnamefont {Y.}~\bibnamefont {Minami}}, \bibinfo
  {author} {\bibfnamefont {M.}~\bibnamefont {Nakajima}}, \ and\ \bibinfo
  {author} {\bibfnamefont {T.}~\bibnamefont {Suemoto}},\ }\href
  {https://journals.aps.org/prl/abstract/10.1103/PhysRevLett.110.137204}
  {\bibfield  {journal} {\bibinfo  {journal} {Phys. Rev. Lett.}\ }\textbf
  {\bibinfo {volume} {110}},\ \bibinfo {pages} {137204} (\bibinfo {year}
  {2013})}\BibitemShut {NoStop}%
\bibitem [{\citenamefont {Mikhaylovskiy}\ \emph {et~al.}(2014)\citenamefont
  {Mikhaylovskiy}, \citenamefont {Hendry}, \citenamefont {Kruglyak},
  \citenamefont {Pisarev}, \citenamefont {Rasing},\ and\ \citenamefont
  {Kimel}}]{THzspectroscopy2}%
  \BibitemOpen
  \bibfield  {author} {\bibinfo {author} {\bibfnamefont {R.~V.}\ \bibnamefont
  {Mikhaylovskiy}}, \bibinfo {author} {\bibfnamefont {E.}~\bibnamefont
  {Hendry}}, \bibinfo {author} {\bibfnamefont {V.~V.}\ \bibnamefont
  {Kruglyak}}, \bibinfo {author} {\bibfnamefont {R.~V.}\ \bibnamefont
  {Pisarev}}, \bibinfo {author} {\bibfnamefont {T.}~\bibnamefont {Rasing}}, \
  and\ \bibinfo {author} {\bibfnamefont {A.~V.}\ \bibnamefont {Kimel}},\ }\href
  {https://journals.aps.org/prb/abstract/10.1103/PhysRevB.90.184405} {\bibfield
   {journal} {\bibinfo  {journal} {Phys. Rev. B}\ }\textbf {\bibinfo {volume}
  {90}},\ \bibinfo {pages} {184405} (\bibinfo {year} {2014})}\BibitemShut
  {NoStop}%
\bibitem [{\citenamefont {Kosen}\ \emph {et~al.}(2018)\citenamefont {Kosen},
  \citenamefont {Morris}, \citenamefont {van Loo},\ and\ \citenamefont
  {Karenowska}}]{magnon10mK}%
  \BibitemOpen
  \bibfield  {author} {\bibinfo {author} {\bibfnamefont {S.}~\bibnamefont
  {Kosen}}, \bibinfo {author} {\bibfnamefont {R.~G.~E.}\ \bibnamefont
  {Morris}}, \bibinfo {author} {\bibfnamefont {A.~F.}\ \bibnamefont {van Loo}},
  \ and\ \bibinfo {author} {\bibfnamefont {A.~D.}\ \bibnamefont {Karenowska}},\
  }\href {https://aip.scitation.org/doi/10.1063/1.5011767} {\bibfield
  {journal} {\bibinfo  {journal} {Appl. Phys. Lett.}\ }\textbf {\bibinfo
  {volume} {112}},\ \bibinfo {pages} {012402} (\bibinfo {year}
  {2018})}\BibitemShut {NoStop}%
\bibitem [{\citenamefont {Tabuchi}\ \emph {et~al.}(2014)\citenamefont
  {Tabuchi}, \citenamefont {Ichino}, \citenamefont {Ishikawa}, \citenamefont
  {Yamazaki}, \citenamefont {Usami},\ and\ \citenamefont {Nakamura}}]{tabuchi}%
  \BibitemOpen
  \bibfield  {author} {\bibinfo {author} {\bibfnamefont {Y.}~\bibnamefont
  {Tabuchi}}, \bibinfo {author} {\bibfnamefont {S.}~\bibnamefont {Ichino}},
  \bibinfo {author} {\bibfnamefont {T.}~\bibnamefont {Ishikawa}}, \bibinfo
  {author} {\bibfnamefont {R.}~\bibnamefont {Yamazaki}}, \bibinfo {author}
  {\bibfnamefont {K.}~\bibnamefont {Usami}}, \ and\ \bibinfo {author}
  {\bibfnamefont {Y.}~\bibnamefont {Nakamura}},\ }\href
  {https://journals.aps.org/prl/abstract/10.1103/PhysRevLett.113.083603}
  {\bibfield  {journal} {\bibinfo  {journal} {Phys. Rev. Lett.}\ }\textbf
  {\bibinfo {volume} {113}},\ \bibinfo {pages} {083603} (\bibinfo {year}
  {2014})}\BibitemShut {NoStop}%
\bibitem [{\citenamefont {Tabuchi}\ \emph {et~al.}(2015)\citenamefont
  {Tabuchi}, \citenamefont {Ichino}, \citenamefont {Noguchi}, \citenamefont
  {Ishikawa}, \citenamefont {Yamazaki}, \citenamefont {Usami},\ and\
  \citenamefont {Nakamura}}]{tabuchiScience}%
  \BibitemOpen
  \bibfield  {author} {\bibinfo {author} {\bibfnamefont {Y.}~\bibnamefont
  {Tabuchi}}, \bibinfo {author} {\bibfnamefont {S.}~\bibnamefont {Ichino}},
  \bibinfo {author} {\bibfnamefont {A.}~\bibnamefont {Noguchi}}, \bibinfo
  {author} {\bibfnamefont {T.}~\bibnamefont {Ishikawa}}, \bibinfo {author}
  {\bibfnamefont {R.}~\bibnamefont {Yamazaki}}, \bibinfo {author}
  {\bibfnamefont {K.}~\bibnamefont {Usami}}, \ and\ \bibinfo {author}
  {\bibfnamefont {Y.}~\bibnamefont {Nakamura}},\ }\href
  {https://science.sciencemag.org/content/349/6246/405} {\bibfield  {journal}
  {\bibinfo  {journal} {Science}\ }\textbf {\bibinfo {volume} {349}},\ \bibinfo
  {pages} {405} (\bibinfo {year} {2015})}\BibitemShut {NoStop}%
\bibitem [{\citenamefont {Prasai}\ \emph {et~al.}(2017)\citenamefont {Prasai},
  \citenamefont {Trump}, \citenamefont {Marcus}, \citenamefont {Akopyan},
  \citenamefont {Huang}, \citenamefont {McQueen},\ and\ \citenamefont
  {Cohn}}]{Tmagnonphonon}%
  \BibitemOpen
  \bibfield  {author} {\bibinfo {author} {\bibfnamefont {N.}~\bibnamefont
  {Prasai}}, \bibinfo {author} {\bibfnamefont {B.~A.}\ \bibnamefont {Trump}},
  \bibinfo {author} {\bibfnamefont {G.~G.}\ \bibnamefont {Marcus}}, \bibinfo
  {author} {\bibfnamefont {A.}~\bibnamefont {Akopyan}}, \bibinfo {author}
  {\bibfnamefont {S.~X.}\ \bibnamefont {Huang}}, \bibinfo {author}
  {\bibfnamefont {T.~M.}\ \bibnamefont {McQueen}}, \ and\ \bibinfo {author}
  {\bibfnamefont {J.~L.}\ \bibnamefont {Cohn}},\ }\href
  {https://journals.aps.org/prb/abstract/10.1103/PhysRevB.95.224407} {\bibfield
   {journal} {\bibinfo  {journal} {Phys. Rev. B}\ }\textbf {\bibinfo {volume}
  {95}},\ \bibinfo {pages} {224407} (\bibinfo {year} {2017})}\BibitemShut
  {NoStop}%
\bibitem [{\citenamefont {Landau}(1932)}]{LZ}%
  \BibitemOpen
  \bibfield  {author} {\bibinfo {author} {\bibfnamefont {L.~D.}\ \bibnamefont
  {Landau}},\ }\href@noop {} {\bibfield  {journal} {\bibinfo  {journal} {Phys.
  Z. Sowjetunion}\ }\textbf {\bibinfo {volume} {2}},\ \bibinfo {pages} {46}
  (\bibinfo {year} {1932})}\BibitemShut {NoStop}%
\bibitem [{\citenamefont {Zener}(1934)}]{LZ2}%
  \BibitemOpen
  \bibfield  {author} {\bibinfo {author} {\bibfnamefont {C.}~\bibnamefont
  {Zener}},\ }\href
  {https://royalsocietypublishing.org/doi/abs/10.1098/rspa.1934.0116}
  {\bibfield  {journal} {\bibinfo  {journal} {Proc. R. Soc. London A}\ }\textbf
  {\bibinfo {volume} {145}},\ \bibinfo {pages} {523} (\bibinfo {year}
  {1934})}\BibitemShut {NoStop}%
\bibitem [{\citenamefont {Bukov}\ \emph {et~al.}(2015)\citenamefont {Bukov},
  \citenamefont {D'Alessio},\ and\ \citenamefont
  {Polkovnikov}}]{FloquetReview}%
  \BibitemOpen
  \bibfield  {author} {\bibinfo {author} {\bibfnamefont {M.}~\bibnamefont
  {Bukov}}, \bibinfo {author} {\bibfnamefont {L.}~\bibnamefont {D'Alessio}}, \
  and\ \bibinfo {author} {\bibfnamefont {A.}~\bibnamefont {Polkovnikov}},\
  }\href {https://www.tandfonline.com/doi/full/10.1080/00018732.2015.1055918}
  {\bibfield  {journal} {\bibinfo  {journal} {Adv. Phys.}\ }\textbf {\bibinfo
  {volume} {64}},\ \bibinfo {pages} {139} (\bibinfo {year} {2015})}\BibitemShut
  {NoStop}%
\bibitem [{\citenamefont {Sato}\ \emph {et~al.}(2016)\citenamefont {Sato},
  \citenamefont {Takayoshi},\ and\ \citenamefont {Oka}}]{STO2016}%
  \BibitemOpen
  \bibfield  {author} {\bibinfo {author} {\bibfnamefont {M.}~\bibnamefont
  {Sato}}, \bibinfo {author} {\bibfnamefont {S.}~\bibnamefont {Takayoshi}}, \
  and\ \bibinfo {author} {\bibfnamefont {T.}~\bibnamefont {Oka}},\ }\href
  {\doibase 10.1103/PhysRevLett.117.147202} {\bibfield  {journal} {\bibinfo
  {journal} {Phys. Rev. Lett.}\ }\textbf {\bibinfo {volume} {117}},\ \bibinfo
  {pages} {147202} (\bibinfo {year} {2016})}\BibitemShut {NoStop}%
\bibitem [{\citenamefont {Nakata}\ \emph {et~al.}(2019)\citenamefont {Nakata},
  \citenamefont {Kim},\ and\ \citenamefont {Takayoshi}}]{KSS}%
  \BibitemOpen
  \bibfield  {author} {\bibinfo {author} {\bibfnamefont {K.}~\bibnamefont
  {Nakata}}, \bibinfo {author} {\bibfnamefont {S.~K.}\ \bibnamefont {Kim}}, \
  and\ \bibinfo {author} {\bibfnamefont {S.}~\bibnamefont {Takayoshi}},\ }\href
  {https://link.aps.org/doi/10.1103/PhysRevB.100.014421} {\bibfield  {journal}
  {\bibinfo  {journal} {Phys. Rev. B}\ }\textbf {\bibinfo {volume} {100}},\
  \bibinfo {pages} {014421} (\bibinfo {year} {2019})}\BibitemShut {NoStop}%
\bibitem [{\citenamefont {Pantazopoulos}\ \emph {et~al.}(2017)\citenamefont
  {Pantazopoulos}, \citenamefont {Stefanou}, \citenamefont {Almpanis},\ and\
  \citenamefont {Papanikolaou}}]{Greece2}%
  \BibitemOpen
  \bibfield  {author} {\bibinfo {author} {\bibfnamefont {P.~A.}\ \bibnamefont
  {Pantazopoulos}}, \bibinfo {author} {\bibfnamefont {N.}~\bibnamefont
  {Stefanou}}, \bibinfo {author} {\bibfnamefont {E.}~\bibnamefont {Almpanis}},
  \ and\ \bibinfo {author} {\bibfnamefont {N.}~\bibnamefont {Papanikolaou}},\
  }\href {\doibase 10.1103/PhysRevB.96.104425} {\bibfield  {journal} {\bibinfo
  {journal} {Phys. Rev. B}\ }\textbf {\bibinfo {volume} {96}},\ \bibinfo
  {pages} {104425} (\bibinfo {year} {2017})}\BibitemShut {NoStop}%
\bibitem [{\citenamefont {Pantazopoulos}\ \emph {et~al.}(2018)\citenamefont
  {Pantazopoulos}, \citenamefont {Papanikolaou},\ and\ \citenamefont
  {Stefanou}}]{Greece4}%
  \BibitemOpen
  \bibfield  {author} {\bibinfo {author} {\bibfnamefont {P.~A.}\ \bibnamefont
  {Pantazopoulos}}, \bibinfo {author} {\bibfnamefont {N.}~\bibnamefont
  {Papanikolaou}}, \ and\ \bibinfo {author} {\bibfnamefont {N.}~\bibnamefont
  {Stefanou}},\ }\href
  {https://iopscience.iop.org/article/10.1088/2040-8986/aaf2c1} {\bibfield
  {journal} {\bibinfo  {journal} {J. Opt}\ }\textbf {\bibinfo {volume} {21}},\
  \bibinfo {pages} {015603} (\bibinfo {year} {2018})}\BibitemShut {NoStop}%
\bibitem [{\citenamefont {Pantazopoulos}\ \emph {et~al.}(2019)\citenamefont
  {Pantazopoulos}, \citenamefont {Tsakmakidis}, \citenamefont {Almpanis},
  \citenamefont {Zouros},\ and\ \citenamefont {Stefanou}}]{Greece5}%
  \BibitemOpen
  \bibfield  {author} {\bibinfo {author} {\bibfnamefont {P.~A.}\ \bibnamefont
  {Pantazopoulos}}, \bibinfo {author} {\bibfnamefont {K.~L.}\ \bibnamefont
  {Tsakmakidis}}, \bibinfo {author} {\bibfnamefont {E.}~\bibnamefont
  {Almpanis}}, \bibinfo {author} {\bibfnamefont {G.~P.}\ \bibnamefont
  {Zouros}}, \ and\ \bibinfo {author} {\bibfnamefont {N.}~\bibnamefont
  {Stefanou}},\ }\href
  {https://iopscience.iop.org/article/10.1088/1367-2630/ab3ad9} {\bibfield
  {journal} {\bibinfo  {journal} {New J. Phys.}\ }\textbf {\bibinfo {volume}
  {21}},\ \bibinfo {pages} {095001} (\bibinfo {year} {2019})}\BibitemShut
  {NoStop}%
\bibitem [{\citenamefont {Pantazopoulos}\ and\ \citenamefont
  {Stefanou}(2019)}]{Greece}%
  \BibitemOpen
  \bibfield  {author} {\bibinfo {author} {\bibfnamefont {P.~A.}\ \bibnamefont
  {Pantazopoulos}}\ and\ \bibinfo {author} {\bibfnamefont {N.}~\bibnamefont
  {Stefanou}},\ }\href {\doibase 10.1103/PhysRevB.99.144415} {\bibfield
  {journal} {\bibinfo  {journal} {Phys. Rev. B}\ }\textbf {\bibinfo {volume}
  {99}},\ \bibinfo {pages} {144415} (\bibinfo {year} {2019})}\BibitemShut
  {NoStop}%
\bibitem [{\citenamefont {Pantazopoulos}\ and\ \citenamefont
  {Stefanou}(2020)}]{Greece3}%
  \BibitemOpen
  \bibfield  {author} {\bibinfo {author} {\bibfnamefont {P.~A.}\ \bibnamefont
  {Pantazopoulos}}\ and\ \bibinfo {author} {\bibfnamefont {N.}~\bibnamefont
  {Stefanou}},\ }\href {\doibase 10.1103/PhysRevB.101.134426} {\bibfield
  {journal} {\bibinfo  {journal} {Phys. Rev. B}\ }\textbf {\bibinfo {volume}
  {101}},\ \bibinfo {pages} {134426} (\bibinfo {year} {2020})}\BibitemShut
  {NoStop}%
\bibitem [{\citenamefont {Nakata}\ \emph {et~al.}(2014)\citenamefont {Nakata},
  \citenamefont {van Hoogdalem}, \citenamefont {Simon},\ and\ \citenamefont
  {Loss}}]{KKPD}%
  \BibitemOpen
  \bibfield  {author} {\bibinfo {author} {\bibfnamefont {K.}~\bibnamefont
  {Nakata}}, \bibinfo {author} {\bibfnamefont {K.~A.}\ \bibnamefont {van
  Hoogdalem}}, \bibinfo {author} {\bibfnamefont {P.}~\bibnamefont {Simon}}, \
  and\ \bibinfo {author} {\bibfnamefont {D.}~\bibnamefont {Loss}},\ }\href
  {https://link.aps.org/doi/10.1103/PhysRevB.90.144419} {\bibfield  {journal}
  {\bibinfo  {journal} {Phys. Rev. B}\ }\textbf {\bibinfo {volume} {90}},\
  \bibinfo {pages} {144419} (\bibinfo {year} {2014})}\BibitemShut {NoStop}%
\bibitem [{\citenamefont {Troncoso}\ and\ \citenamefont
  {Nunez}(2014)}]{Troncoso}%
  \BibitemOpen
  \bibfield  {author} {\bibinfo {author} {\bibfnamefont {R.~E.}\ \bibnamefont
  {Troncoso}}\ and\ \bibinfo {author} {\bibfnamefont {A.~S.}\ \bibnamefont
  {Nunez}},\ }\href
  {https://www.sciencedirect.com/science/article/abs/pii/S0003491614000980?via%3Dihub}
  {\bibfield  {journal} {\bibinfo  {journal} {Ann. Phys.}\ }\textbf {\bibinfo
  {volume} {346}},\ \bibinfo {pages} {182} (\bibinfo {year}
  {2014})}\BibitemShut {NoStop}%
\bibitem [{\citenamefont {Nikuni}\ \emph {et~al.}(2000)\citenamefont {Nikuni},
  \citenamefont {Oshikawa}, \citenamefont {Oosawa},\ and\ \citenamefont
  {Tanaka}}]{oshikawa}%
  \BibitemOpen
  \bibfield  {author} {\bibinfo {author} {\bibfnamefont {T.}~\bibnamefont
  {Nikuni}}, \bibinfo {author} {\bibfnamefont {M.}~\bibnamefont {Oshikawa}},
  \bibinfo {author} {\bibfnamefont {A.}~\bibnamefont {Oosawa}}, \ and\ \bibinfo
  {author} {\bibfnamefont {H.}~\bibnamefont {Tanaka}},\ }\href
  {https://link.aps.org/doi/10.1103/PhysRevLett.84.5868} {\bibfield  {journal}
  {\bibinfo  {journal} {Phys. Rev. Lett.}\ }\textbf {\bibinfo {volume} {84}},\
  \bibinfo {pages} {5868} (\bibinfo {year} {2000})}\BibitemShut {NoStop}%
\bibitem [{\citenamefont {Ueda}\ and\ \citenamefont
  {Totsuka}(2009)}]{TotsukaBEC2}%
  \BibitemOpen
  \bibfield  {author} {\bibinfo {author} {\bibfnamefont {H.~T.}\ \bibnamefont
  {Ueda}}\ and\ \bibinfo {author} {\bibfnamefont {K.}~\bibnamefont {Totsuka}},\
  }\href {https://link.aps.org/doi/10.1103/PhysRevB.80.014417} {\bibfield
  {journal} {\bibinfo  {journal} {Phys. Rev. B}\ }\textbf {\bibinfo {volume}
  {80}},\ \bibinfo {pages} {014417} (\bibinfo {year} {2009})}\BibitemShut
  {NoStop}%
\bibitem [{\citenamefont {Giamarchi}\ \emph {et~al.}(2008)\citenamefont
  {Giamarchi}, \citenamefont {R{\"u}egg},\ and\ \citenamefont
  {Tchernyshyov}}]{Giamarchi2008NatPhys}%
  \BibitemOpen
  \bibfield  {author} {\bibinfo {author} {\bibfnamefont {T.}~\bibnamefont
  {Giamarchi}}, \bibinfo {author} {\bibfnamefont {C.}~\bibnamefont
  {R{\"u}egg}}, \ and\ \bibinfo {author} {\bibfnamefont {O.}~\bibnamefont
  {Tchernyshyov}},\ }\href {https://doi.org/10.1038/nphys893} {\bibfield
  {journal} {\bibinfo  {journal} {Nat. Phys.}\ }\textbf {\bibinfo {volume}
  {4}},\ \bibinfo {pages} {198} (\bibinfo {year} {2008})}\BibitemShut {NoStop}%
\end{thebibliography}
\end{document}